\documentclass[ALICE,manyauthors]{cernphprep}
\pdfoutput=1
\RequirePackage{ifpdf}

\usepackage{graphicx}
\usepackage{amsmath}
\usepackage{amssymb}
\usepackage{xspace}
\usepackage{float}
\usepackage{placeins}
\usepackage{lineno}
\usepackage[pdftex,usenames,dvipsnames]{xcolor}

\ifpdf
  \usepackage[pdftex]{hyperref}

  \hypersetup{
    final,
    a4paper=true,
    colorlinks=false,
    bookmarks=true,
    pdffitwindow=true,
    pdfnewwindow=true,	
    pdfauthor={ALICE Collaboration},
    pdfcreator={pdflatex},
    pdfproducer={ALICE Collaboration},
    pdftitle={Particle Identification in ALICE: a Bayesian approach},
    pdfsubject={Particle Identification in ALICE: a Bayesian approach},
    pdfkeywords={}
  }

\else
  \usepackage[pdftex]{hyperref}
\fi

\newcommand {\pT}        {\pt}
\newcommand {\meanpT}    {\ensuremath{\langle p_{\mathrm{T}} \kern-0.1em\rangle}\xspace}
\newcommand {\mean}[1]   {\ensuremath{\langle #1 \kern-0.1em\rangle}\xspace} 
\newcommand {\sqrtsNN}   {\ensuremath{\sqrt{s_{\textsc{NN}}}}\xspace}
\newcommand {\sqrts}     {\ensuremath{\sqrt{s}}\xspace}

\newcommand {\cTau}      {\ensuremath{c\tau}}

\newcommand{\dEdx}       {\ensuremath{\mathrm{d}E/\mathrm{d}x}\xspace}
\newcommand{\ttof}       {\ensuremath{t_\mathrm{TOF}}\xspace}
\newcommand {\pp}        {\mbox{$\mathrm {p\kern-0.05em p}$}\xspace}
\newcommand {\ppBoldMath} {\mbox{$\mathrm { \mathbf p\kern-0.05em \mathbf p }$}\xspace}

\newcommand {\PbPb}      {\ensuremath{\mbox{Pb--Pb}}\xspace}

\newcommand {\pPb}       {\ensuremath{\mbox{p--Pb}}\xspace}

\newcommand {\MeanNpart} {\mbox{\ensuremath{< \kern-0.15em N_{part} \kern-0.15em >}}}

\newcommand {\sig}       {\ensuremath{S}\xspace}
\newcommand {\expsig}    {\ensuremath{\hat{S}}\xspace}
\newcommand {\prob}      {\ensuremath{P}\xspace}
\newcommand {\prior}     {\ensuremath{C}\xspace}
\newcommand {\prop}      {\ensuremath{F}\xspace}
\newcommand {\atrue}     {\ensuremath{\vec{A}_{\mathrm{true}}}\xspace}
\newcommand {\ameas}     {\ensuremath{\vec{A}_{\mathrm{meas}}}\xspace}
\newcommand {\detresp}   {\ensuremath{R}\xspace}

\newcommand {\pid}       {\ensuremath{\mathrm{\epsilon}_\mathrm{PID}}\xspace}
\newcommand {\nsigma}    {\ensuremath{\mathrm{n_{\sigma}}}\xspace}

\newcommand {\mass}     {\mbox{\rm MeV$\kern-0.15em /\kern-0.12em c^2$}}
\newcommand {\tev}      {\mbox{${\rm TeV}$}\xspace}

\newcommand {\mmom}     {\mbox{\rm MeV$\kern-0.15em /\kern-0.12em c$}}
\newcommand {\gmom}     {\mbox{\rm GeV$\kern-0.15em /\kern-0.12em c$}}
\newcommand {\mmass}    {\mbox{\rm MeV$\kern-0.15em /\kern-0.12em c^2$}}
\newcommand {\gmass}    {\mbox{\rm GeV$\kern-0.15em /\kern-0.12em c^2$}}

\newcommand {\mim}      {\mbox{$ \mu {\rm m}$}}

\newcommand {\dg}       {\mbox{$\kern+0.1em ^\circ$}}

\newcommand{\piMinus}           {\ensuremath{\mathrm {\pi^-}}\xspace}
\newcommand{\piPlus}            {\ensuremath{\mathrm {\pi^+}}\xspace}

\newcommand{\proton}    {\mbox{$\mathrm {p}$}\xspace}

\newcommand{\DZero}     {\mbox{$\mathrm {D^0}$}\xspace}

\newcommand{\Dmes}      {\mbox{$\mathrm {D}$}\xspace}
\newcommand{\Lc}        {\mbox{$\mathrm {\Lambda_{c}^{+}}$}\xspace}

\newcommand{\rmLambda}          {\mbox{$\mathrm {\Lambda}$}\xspace}

\newcommand{\rmLambdas}         {\mbox{$\mathrm {\Lambda \kern-0.2em + \kern-0.2em \overline{\Lambda}}$}\xspace}

\newcommand{\Vzero}             {\mbox{$\mathrm {V^0}$}\xspace}

\newcommand{\Kzs}               {\ensuremath{\mathrm {K^0_S}}\xspace}
\newcommand{\phimes}            {\ensuremath{\mathrm {\phi}}\xspace}
\newcommand{\Kminus}            {\ensuremath{\mathrm {K^-}}\xspace}
\newcommand{\Kplus}             {\ensuremath{\mathrm {K^+}}\xspace}
\newcommand{\Kstar}             {\mbox{$\mathrm {K^*}$}\xspace}

\newcommand{\TeV}{\ensuremath{\mathrm{TeV}}\xspace}
\newcommand{\gevc}{\ensuremath{\mathrm{GeV}/c}\xspace}
\newcommand{\GeVc}{\gevc}

\newcommand{\pt}{\ensuremath{p_{\rm T}}\xspace}
\newcommand{\DtoKpi}{\ensuremath{\rm D^0\to K^-\pi^+}\xspace}

\newcommand{\Dzero}{\ensuremath{\mathrm {D^0}}\xspace}
\newcommand{\Dzerobar}{\ensuremath{\mathrm{\overline{D}^0}}\xspace}

\newcommand{\LctopKpi}{\ensuremath{\rm \Lambda_{c}^{+}\to p K^-\pi^+}\xspace}

\newcommand{\run}[1]{\textsc{Run\,#1}}                                                                                                                                                                     

\newcommand{\figref}[1]{Fig.~\ref{#1}}
\newcommand{\figrefs}[1]{Figs.~\ref{#1}}
\newcommand{\Figref}[1]{Figure~\ref{#1}}

\newcommand{\secref}[1]{Section~\ref{#1}}
\newcommand{\Secref}[1]{Section~\ref{#1}}
\begin{document}
\begin{titlepage}
\PHyear{2016}
\PHnumber{023}
\PHdate{02 February}


\title{Particle identification in ALICE: a Bayesian approach}

\Collaboration{ALICE Collaboration\thanks{See Appendix~\ref{app:collab} for the list of collaboration members}}
\ShortAuthor{ALICE Collaboration} 

%
%
%
\date{Received: date / Revised version: date}
%
\begin{abstract}
We present a Bayesian approach to particle identification (PID) within the ALICE experiment. The aim is to more effectively combine
the particle identification capabilities of its various detectors.
After a brief explanation of the adopted methodology and formalism, the performance of the Bayesian PID approach for charged pions, kaons and protons in the central barrel of ALICE is studied. PID is performed via measurements of specific energy loss ($\mathrm{d}E/\mathrm{d}x$) and time-of-flight. PID efficiencies and misidentification probabilities are extracted and compared with Monte Carlo simulations using high-purity samples of identified particles in the decay channels $\Kzs \rightarrow \piMinus\piPlus$, $\phimes \rightarrow \Kminus\Kplus$, and 
$\rmLambda \rightarrow \proton\piMinus$ in \pPb collisions at $\sqrtsNN=5.02\,\TeV$.
In order to thoroughly assess the validity of the Bayesian approach, this methodology was used to obtain corrected $\pt$ spectra of pions, kaons, protons, and \Dzero mesons in pp collisions at $\sqrts=7$\,\TeV. In all cases, the results using Bayesian PID were found to be consistent with previous measurements performed by ALICE using a standard PID approach. For the measurement of \DtoKpi, it was found that a Bayesian PID approach gave a higher signal-to-background ratio and a similar or larger statistical significance when compared with standard PID selections, despite a reduced identification efficiency. 
Finally, we present an exploratory study of the measurement of \LctopKpi in pp collisions at $\sqrts=7$\,\TeV, using the Bayesian approach for the identification of its decay products.

\end{abstract}
\end{titlepage}
\setcounter{page}{2}

\maketitle
%
\section{Introduction}
\label{intro}

Particle Identification (PID) provides information about the mass and flavour composition of particle production in high-energy physics experiments. In the context of ALICE (A Large Ion Collider Experiment)~\cite{aliceperf}, identified particle yields and spectra give access to the properties of the state of matter formed at extremely high energy densities in ultra-relativistic heavy-ion collisions. Modern experiments usually consist of a variety of detectors featuring different PID techniques.
Bayesian approaches to the problem of combining PID signals from different detectors have already been used by several experiments, e.g. NA27~\cite{pidNA27}, HADES~\cite{pidHADES} and BESII~\cite{pidBESII}. 
This technique was proposed for ALICE during its early planning stages~\cite{CHEP04} and then used extensively to prepare the ALICE Physics Performance Report~\cite{ppr1,ppr2}. An analogous method is also used to combine the PID signals from the different layers of the ALICE Transition Radiation Detector.

The ALICE detector system is composed of a central part that covers the mid-rapidity region $|\eta|<1$
(the `central barrel'), and a muon spectrometer that covers the forward rapidity region
$-4<\eta<-2.5$. 
The central barrel detectors that have full coverage in azimuth ($\varphi$) are, from small to large radii, the
Inner Tracking System (ITS), the Time Projection Chamber (TPC), the Transition Radiation 
Detector (TRD)\footnote{The TRD was completed during the first Long Shutdown phase (2013--2015) and only had partial $\varphi$
coverage during the \run{1} data taking period (2009--2013).}
and the Time Of Flight system (TOF).
Further dedicated PID detectors with limited acceptance in $\varphi$ and $\eta$ are also located in the central barrel.
These are the Electromagnetic Calorimeter (EMCal), the Photon Spectrometer (PHOS), and a Cherenkov system for 
High-Momentum Particle Identification (HMPID).

The central barrel detectors provide complementary PID information and the capability to separate particle species in different momentum intervals.
At low momenta ($p\lesssim 3$--$4\,\gevc$), a track-by-track separation of pions, kaons and protons is made possible by combining the PID signals from different detectors.
At higher momenta, statistical unfolding based on the relativistic rise of the TPC signal can be performed for PID. 
Given the wide range of momenta covered, ALICE has the strongest PID capabilities of any of the LHC experiments.
More details about the identification possibilities of the single detectors can be found in~\cite{aliceperf}.

Several different PID methods were applied for analyses of 
data collected by ALICE during the \run{1} data taking period of the LHC (2009--2013).
A non-exhaustive list of examples of PID in ALICE is given in the following. Results were published on the transverse momentum (\pt) distributions of charged pion, kaon and proton production~\cite{spectra7TeV,spectra2.76TeV,spectraPbPb} in
different collision systems and centre-of-mass energies using the ITS, TPC, TOF and HMPID detectors. 
Electron measurements from semileptonic heavy-flavour hadron decays took advantage of the TPC, TOF, TRD and EMCal detectors~\cite{HFE1,HFE2}. 
Neutral pion production was studied via photon detection in the PHOS and the detection of $\rm e^+e^-$ pairs from gamma conversions in the TPC.
 PID detectors were also used extensively to improve the signal-to-background ratios when studying the production of certain particles based on the reconstruction of their decay products, such as \Dmes mesons~\cite{alice2011charmDmesonsInPP7TeV}, \phimes and \Kstar resonances~\cite{Phi}; in studies of particle correlations, such as femtoscopy~\cite{femto}; and to identify light nuclei~\cite{nucleiMass}. In all analyses where PID was used,
selections were applied based on individual detector signals and later combined. 

In this paper we describe results obtained during \run{1} in \pp collisions at $\sqrts=7\,\tev$, \PbPb collisions at $\sqrtsNN=2.76\,\tev$ and \pPb collisions at $\sqrtsNN=5.02\,\tev$.
In particular, we focus on the hadron identification capabilities of the central barrel detectors
that had full azimuthal coverage during \run{1}.
These are the ITS, TPC and TOF detectors, which are described in more detail below.
The other central barrel detectors are not discussed here.

The ITS is formed of six concentric cylindrical layers of silicon detectors: two layers each of Silicon Pixel (SPD), 
Silicon Drift (SDD) and Silicon Strip Detectors (SSD).
The SDD and the SSD provide a read-out of the signal amplitude, and thus contribute to the PID by measuring the specific energy loss (\dEdx)
of the traversing particle. 
A truncated mean of the (up to) four signals is calculated, resulting in a relative \dEdx resolution of 
approximately 12\%~\cite{aliceperf}.

The TPC~\cite{TDR:tpc} is the main tracking device of ALICE.
Particle identification is performed by measuring the specific energy loss (\dEdx) in the detector gas\footnote{
A Ne-CO$_2$-based gas mixture of $\sim$90$\,\mathrm{m}^3$ in \run{1}.}
in up to 159 read-out pad rows.
A truncated mean that rejects the 40\% largest cluster charges is built, resulting in a Gaussian \dEdx response.
The \dEdx resolution ranges between 5--8\% depending on the track inclination angle and drift distance,
the energy loss itself, and the centrality in \pPb and \PbPb collisions due to the differing detector occupancy.

The TOF detector~\cite{TOFperf} is based on Multigap Resistive Plate Chamber technology.
It measures the flight times of particles with an intrinsic resolution of $\sim$80\,ps.
The expected flight time for each particle species is calculated during the reconstruction, and then PID
is performed via a comparison between the measured and expected times.

Other forward-rapidity detectors are also relevant to the data analyses presented in this paper.
The V0 plastic scintillator arrays, V0A (covering $2.8 < \eta < 5.1$) and V0C ($-3.7 < \eta < -1.7$), are required in the minimum-bias trigger. The V0 signals are also used to determine the centrality of \pPb and \PbPb collisions.
The T0 detector is a quartz Cherenkov detector that is used for start-time estimation
with a resolution of $\sim$40\,ps in pp collisions and 20--25\,ps in \PbPb~collisions.

A thorough understanding of the detector response is crucial for any particle identification method.
In the case of \dEdx, this means a well parameterised description of the average energy loss (according to the Bethe-Bloch formula~\cite{blumrolandi}) and a reliable estimate of the signal resolution.
For the TOF measurement, the start-time information, start-time resolution, track reconstruction resolution, and intrinsic detector resolution need to be known~\cite{TOFperf}.

Combining the PID signals of the individual detectors using a Bayesian approach~\cite{Bayes} makes effective use of the full PID capabilities of ALICE. However, using combined probabilities to perform particle identification may result in unintuitive,
non-trivial track selections. 
It is therefore important to benchmark the Bayesian PID method, compare efficiencies in data and Monte Carlo, and validate that this technique does not introduce a systematic bias with respect to previously published results. This paper focuses on the verification of the Bayesian PID approach on the basis of various data analyses.

\Secref{sec:pidmethods} describes the Bayesian approach (\ref{sec:bayesapp}), the definitions of efficiency and contamination in the context of PID (\ref{sec:effcontam}), the extraction and application of prior distributions (\ref{sec:priors}), and
different strategies for using the resulting probabilities (\ref{sec:strategies}).
\Secref{sec:V0s} presents benchmark analyses of high-purity samples of pions, kaons, and protons from the two-prong decays of \Kzs mesons, \phimes mesons and $\Lambda$ baryons.
\Secref{results} presents validations of the Bayesian approach for two full analyses:
the measurement of the transverse momentum spectra of pions, kaons
and protons (\ref{sec:spectra}), and the analysis of $\DtoKpi$ (\ref{sec:d0res}).
\Secref{sec:lcres} illustrates the application of the Bayesian PID approach to maximise the statistical significance when analysing the production of $\Lc$ baryons.
Finally, a conclusion and outlook are given in \secref{concl}. 

\section{Bayesian PID in ALICE}\label{sec:pidmethods}

\begin{figure}[t]
  \centering
  \includegraphics[width=0.6\textwidth]{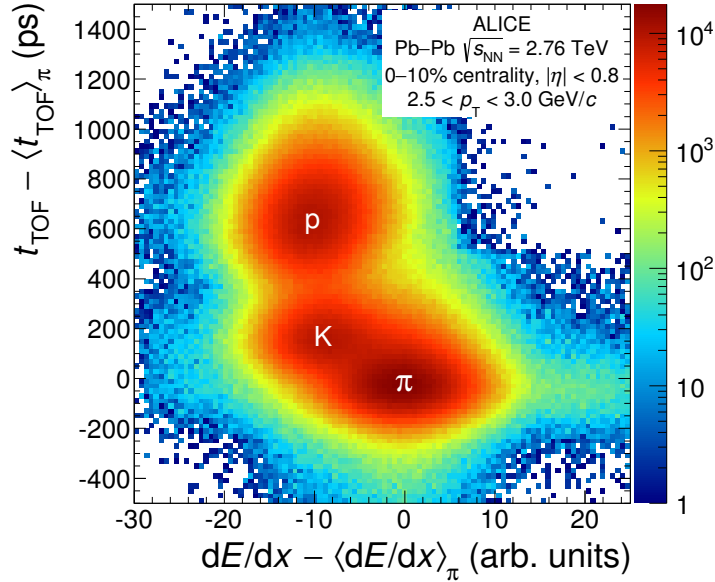}
  \caption[Combined TPC-TOF PID.]
          {Combined particle identification in the TPC and TOF for data from \PbPb collisions at $\sqrtsNN=2.76\,\TeV$, shown as a two-dimensional plot.
           The PID signals are expressed in terms of the deviation
           from the expected response for pions in each detector.}
  \label{fig:2Dpid}
\end{figure}

Simple selections based on the individual PID signals of each detector do not take full advantage of the PID capabilities of ALICE.
An example of this is illustrated in \figref{fig:2Dpid}, which shows the separation of the expected TPC and TOF signals for pions, kaons and protons with transverse momenta in the range $2.5<\pt<3$\,\GeVc.
Clearly, the separation in the two-dimensional plane (the peak-to-peak distance) of e.g. pions and kaons is larger than the
separation of each individual one-dimensional projection.
A natural way of combining the information of independent detectors is to express the signals in
terms of probabilities.
An additional advantage of this method is that detectors with non-Gaussian responses can also be
included in a straightforward way.
A Bayesian approach makes use of the full PID capabilities by folding the probabilities with 
the expected abundances (priors) of each particle species.
This section outlines the standard `\nsigma' PID approach, before presenting the method used to combine the signals from the different detectors when adopting a Bayesian PID approach.

\subsection{PID signals in ALICE}\label{sec:bayesapp}

The response of each detector can be expressed in terms of its raw signal, \sig.
One of the simplest ways of performing PID is to directly select based on \sig. Examples of \sig include the flight-time information from the TOF detector, \ttof, and the specific energy loss \dEdx in detector gas or silicon, respectively measured by the TPC and the ITS.
A more advantageous approach would be to use a discriminating variable, $\xi$, which makes use of the expected detector response \detresp:
\begin{equation}
  \xi = f(\sig, \detresp),
\end{equation}
where \detresp can have functional dependences on the properties of the particle tracks, typically the momentum $p$, the charge $Z$, or the track length $L$. 
The detector response functions are usually complex parameterisations expressing an in-depth knowledge of subtle detector effects. 

For a detector with a Gaussian response, \detresp is given by the expected average signal $\expsig(H_i)$ for a given particle species $H_i$  and the expected signal resolution $\sigma$.
The index $i$ usually refers to electrons, muons, pions, kaons or protons, but may also include light nuclei such as deuterons, 
tritons, $^3\mathrm{He}$ nuclei and $^4\mathrm{He}$ nuclei.
The most commonly used discriminating variable for PID is the \nsigma variable, defined as the deviation of the measured signal from that expected for a species $H_i$, in terms of the detector resolution:
\begin{equation}
  \mathrm{n}_{\sigma_{\scriptstyle\alpha}^{\scriptstyle i}} = \frac{ \sig_{\alpha} - \expsig(H_i)_{\alpha} }{\sigma_{\alpha}^i},
\end{equation}
where $\rm \alpha=(ITS, TPC, ...)$. The resolution is given here as $\sigma_\alpha^i$, as it depends both on the detector and the species being measured. In the following, $\sigma_\alpha^i$ is simply referred to as `$\sigma$'.

The \nsigma PID approach corresponds to a `true/false' decision on whether a particle belongs to a given species. A certain identity is assigned to a track if this value
lies within a certain range around  the expectation (typically 2 or 3$\sigma$). Depending on
the detector separation power, a track can be compatible with more than one identity.

For a given detector $\alpha$ with a Gaussian response, it is possible to define the conditional probability that a particle of species $H_i$ will produce a signal \sig as
\begin{equation}\label{eq:detWeight}
  \prob(\sig|H_i) = \frac{1}{\sqrt{2\pi}\sigma} e^{-\frac{1}{2}\mathrm{n_{\sigma}}^2} =
  \frac{1}{\sqrt{2\pi}\sigma} e^{-\frac{( \sig - \expsig(H_i) )^2}{2\sigma^2}}.
\end{equation}

In the case of a non-Gaussian response, the probability is described by an alternative parameterisation appropriate to the detector.
The advantage of using probabilities is that the probabilities from different detectors, $\prob_{\alpha}$, with and without Gaussian responses, can then be combined as a product:
\begin{equation}
  \prob(\vec{\sig}|H_i) = \prod_{\alpha=\mathrm{ITS, TPC, ...}} \prob_{\alpha}(\sig_{\alpha}|H_i),
\end{equation}
where $\vec{\sig}=(\sig_{\mathrm{ITS}},\sig_{\mathrm{TPC}},...)$.

The probability estimate $\prob(\vec\sig|H_i)$ (either for a single detector, or combined over many) can be interpreted as the conditional probability 
that the set of detector signals $\vec\sig$ will be seen for a given particle species $H_i$.
However, the variable of interest is the conditional probability that the particle is of species $H_i$, given some measured detector signal (i.e. $\prob(H_i|\vec\sig)$).
The relation between the two for a combined set of detectors can be expressed using Bayes' theorem~\cite{Bayes}:
\begin{equation}\label{eq:BayesEq}
  \prob(H_i|\vec{\sig}) = \frac{\prob(\vec{\sig}|H_i)\prior(H_i)}{\sum_{k=e,\mu,\pi,...}\prob(\vec{\sig}|H_k) \prior(H_k)}.
\end{equation}

Here, $\prior(H_i)$ is an {\it a priori} probability of measuring the particle species $H_i$, also known as the \emph{prior},
and the conditional probability $\prob(H_i|\vec{\sig})$ is known as the \emph{posterior} probability.

The priors (which are discussed in more detail in~\secref{sec:priors}) serve as a `best guess' of the true particle yields per event. When such a definition is adopted for the priors, a selection based on the
Bayesian probability calculated with Eq.~\ref{eq:BayesEq} then corresponds to a request on the purity 
(defined as the ratio between the number of correctly identified particles and the total selected). 
Additionally, priors can be used to reject certain particle species that are not relevant to a given analysis.
Most commonly in the context of the ALICE central barrel, the prior for muons is set to zero.
Due to the similarity between the pion and muon mass, the two species are almost indistinguishable over a broad momentum range; the efficiency of detecting a pion would thus be reduced if muons were not neglected. At the same time, this influences the number of particles wrongly identified as pions, since 
the true abundance of muons (roughly 2\% of all particles, estimated using Monte Carlo simulations) is neglected. This case is further discussed in the following sections.

\subsection{PID efficiency and contamination}\label{sec:effcontam}

In order to obtain the physical quantity
of interest (typically a cross section or a spectrum) from a raw yield, it is necessary to (a) compute the efficiency due to other selections applied before PID, and (b) compute the efficiency of the PID strategy. The PID efficiency is defined as the proportion of particles of a given species that are identified correctly by the PID selections.
Both kinds of efficiency are usually estimated via Monte Carlo techniques. To precisely compute the efficiency of a given PID
strategy, it is of utmost importance that an accurate description of the actual signals
present in the data is provided by the Monte Carlo simulation.
Special care must be given to the `tuning' of Monte Carlo simulations to reproduce all of the features and dependences
observed in data for PID signals.

It is possible to define a \pid matrix that contains the probability to identify a species \emph{i}
as a species \emph{j}. If only pions, kaons and protons are considered, the $3 \times 3$ \pid matrix is defined as

\begin{equation}\label{eq:pidMatrix}
\pid = \begin{pmatrix}
\epsilon_{\pi\pi} & \epsilon_{\pi \mathrm{K}} & \epsilon_{\pi \mathrm{p}} \\
\epsilon_{\mathrm{K}\pi} & \epsilon_{\mathrm{K}\mathrm{K}}  & \epsilon_{\mathrm{Kp}} \\
\epsilon_{\mathrm{p}\pi} & \epsilon_{\mathrm{p}\mathrm{K}} & \epsilon_\mathrm{pp} 
\end{pmatrix},
\end{equation}
where the diagonal elements  $\epsilon_{ii}$ are the PID efficiencies,
and the non-diagonal elements ($\epsilon_{ij}$, $i\ne j$) represent the probability of misidentifying a species \emph{i} as a different species \emph{j}. 

The abundance vectors for pions, kaons and protons are defined as
\begin{equation}\label{eq:Avectordef}
\ameas =   \begin{pmatrix}
      \pi_{\mathrm{meas}} \\
      \mathrm{K}_{\mathrm{meas}} \\
      \mathrm{p}_{\mathrm{meas}}
    \end{pmatrix}{\text{~and~~}} \atrue =    \begin{pmatrix}
      \pi_{\mathrm{true}} \\
      \mathrm{K}_{\mathrm{true}} \\
      \mathrm{p}_{\mathrm{true}}
    \end{pmatrix},
\end{equation}
\noindent where the elements of \ameas $(\atrue)$ represent the measured (true) abundances of each species.
The diagonal elements $\epsilon_{ii}$ of the PID matrix are then defined as

\begin{equation}
\epsilon_{ii} = \frac{N_{i \text{ identified as }i}}{A_{\text{true}}^i}.
\end{equation}

Techniques to estimate the matrix elements 
are discussed in \secref{sec:V0s}. The abundance vectors \ameas and \atrue are linked by the following relation:

\begin{equation}\label{eq:abundances}
    \begin{pmatrix}
      \pi_{\mathrm{meas}} \\
      \mathrm{K}_{\mathrm{meas}} \\
      \mathrm{p}_{\mathrm{meas}}
    \end{pmatrix}
    = 
    \begin{pmatrix}
\epsilon_{\pi\pi} & \epsilon_{\pi \mathrm{K}} & \epsilon_{\pi \mathrm{\mathrm{p}}} \\
\epsilon_{\mathrm{K}\pi} & \epsilon_{\mathrm{K}\mathrm{K}}  & \epsilon_{\mathrm{K}\mathrm{\mathrm{p}}} \\
\epsilon_{\mathrm{p}\pi} & \epsilon_{\mathrm{p}\mathrm{K}} & \epsilon_{\mathrm{p}\mathrm{\mathrm{p}}} 
    \end{pmatrix}^\intercal
    \cdot
    \begin{pmatrix}
      \pi_{\mathrm{true}} \\
      \mathrm{K}_{\mathrm{true}} \\
      \mathrm{p}_{\mathrm{true}}
    \end{pmatrix}.
  \end{equation}

Inverting the $\pid^{\intercal}$ matrix, the physical quantities are then extracted via:

\begin{equation}\label{eq:trueabund}
\atrue = (\pid^{\intercal})^{-1} \times \ameas.
\end{equation}

The \pid matrix elements have functional dependences on many variables, primarily $\pt$ and collision system. Other second-order dependences (for example pseudorapidity,
event multiplicity, and centrality) can also be studied depending on the specific track selections and PID strategies applied.

The contamination of the species \emph{j} due to a different species \emph{i} $(c_{ji})$ is the number of particles belonging
to species \emph{i} that are wrongly identified as \emph{j} $(N_{i~\text{identified as}~j})$, divided by the total number of identified \emph{j} particles $(A_\mathrm{meas}^j)$~\cite{ppr2}:

\begin{equation}\label{eq:contamination}
  c_{ji} = \frac{N_{i~\text{identified as }j}}{A_\mathrm{meas}^j}, i \ne j.
\end{equation}

Contamination should not be confused with the misidentification probabilities defined in Eq.~\ref{eq:pidMatrix}, which do not
depend on the real abundances. The connection between contamination and misidentification is:

\begin{equation}\label{eq:misProb}
  c_{ji} = \frac{\epsilon_{ij} A_{\mathrm{true}}^{i}}{\epsilon_{jj}A_{\mathrm{true}}^{j} + \sum_{j\ne k}{\epsilon_{jk}A_{\mathrm{true}}^{k}}}.
\end{equation}

There is usually a trade-off between efficiency and contamination. Both depend on
the PID strategy (e.g. choice of detectors), the detector response, and the priors used, all of which are momentum-dependent. The contamination is additionally 
driven by the real abundances.

An accurate estimate of the contamination $c_{ji}$ 
depends on the real abundances, and it must be determined with data-driven techniques (or with Monte Carlo simulations with abundances 
corresponding to what is found in the data).

Although the PID matrix elements are independent of the real abundances, they still depend on the choice of priors. They are therefore evaluated consistently as long as the same set of priors is used both in the data analysis and for the Monte Carlo simulations, provided that the detector responses are simulated correctly.
In addition, a choice of priors that lies closer to the true abundances allows the best compromise to be found between the maximisation of the efficiency and the minimisation of the contamination probabilities. Considering that any systematic uncertainties in the detector response will be amplified if the efficiency is low or if the contamination probabilities are large, the method becomes more effective as the priors tend closer to reality.
An extreme example of this would be the identification of pions and muons using equal priors for all species, as the two species are difficult to distinguish from each other ($\varepsilon_{\pi\pi} \approx \varepsilon_{\mu\mu} \approx 0.5$ and $\varepsilon_{\pi\mu} \approx \varepsilon_{\mu\pi} \approx 0.5$). In such a case, even a small discrepancy in the description of the detector response in Monte Carlo would cause a fluctuation in $\varepsilon_{ij}$, leading to a large uncertainty in the estimate of the pion yield; a choice of priors corresponding to the true abundances would prevent this from happening.

In cases where the detector responses are well separated between different species, the priors do not need to correspond to the true abundances, and could even be flat. The differences between different choices can then be used to provide an estimate of the systematic uncertainties depending on the current knowledge of the detector response. The influence of the choice of priors is further discussed as part of the analyses presented in~\secref{results}.

The combination of probabilities can also be used to identify and reduce the level of unphysical background that may arise in PID analyses due
to track misassociation between detectors. 
For example, the fraction of TPC tracks that are not correctly associated with the corresponding TOF hit increases with the multiplicity of the event, and depends
on the spatial matching window used in the reconstruction to associate a TOF hit to a track.
In ALICE this effect only plays a role in \PbPb and \pPb collisions, and even in the most central \PbPb collisions it remains below 10\% for tracks with momentum above 1\,\GeVc.
In these cases, a mismatch probability can be defined based on e.g. the measured 
flight time being uncorrelated with the track reconstructed with the ITS and TPC due to the TOF hit being produced by  another particle.

\subsection{Priors in ALICE}\label{sec:priors}

As discussed in \secref{sec:bayesapp}, an analysis using the Bayesian approach is expected
to be moderately dependent on the choice of priors. Separate sets of priors were evaluated for each collision system.
They were evaluated by means of an iterative procedure on data taken in 2010 and 2013 from \pp, \PbPb, and \pPb collisions at $\sqrts=7\,\tev$, $\sqrtsNN=2.76\,\tev$ and $\sqrtsNN=5.02$\,\tev, respectively, and were computed
as a function of transverse momentum. Priors were also determined as a function of centrality for \PbPb collisions, and of the multiplicity class (based on the signal in the V0A detector) for \pPb~data. All of the priors were obtained at mid-rapidity, $|\eta|<0.8$. 

The absolute normalisation of the priors is arbitrary, and was chosen so as to normalise all of the priors to the abundance of pions. The value of the priors for pions is thereby set to unity for all \pt.
Flat priors (i.e. 1 for all species) are applied at the beginning of the iterative procedure.
Bayesian posterior probabilities $P_n(H_i|S)$ are computed using the priors obtained in step $n$, as defined in Eq.~\ref{eq:BayesEq}. These probabilities are then used in turn as weights to fill identified \pT~spectra $Y(H_i,p_{\rm T})$ for the step $n+1$ starting
from the inclusive (unidentified) measured \pT spectra:
\begin{equation}\label{eq:loopYields}
Y_{n+1}(H_i,p_{\rm T})=\sum_{S}P_n(H_i|S),
\end{equation}
where the summation is performed for all signals $S$ induced by particles of a given \pT in the sample.
It is then possible to obtain a new set of priors $C_{n+1}$ from the relative ratios of the identified spectra according to

\begin{equation}\label{eq:loopPriors}
C_{n+1}(H_i,p_{\rm T})=\frac{Y_{n+1}(H_i,p_{\rm T})}{Y_{n+1}(H_\pi,p_{\rm T})}.
\end{equation}

\begin{figure}[htb]
\centering
\includegraphics[width=0.95\textwidth]{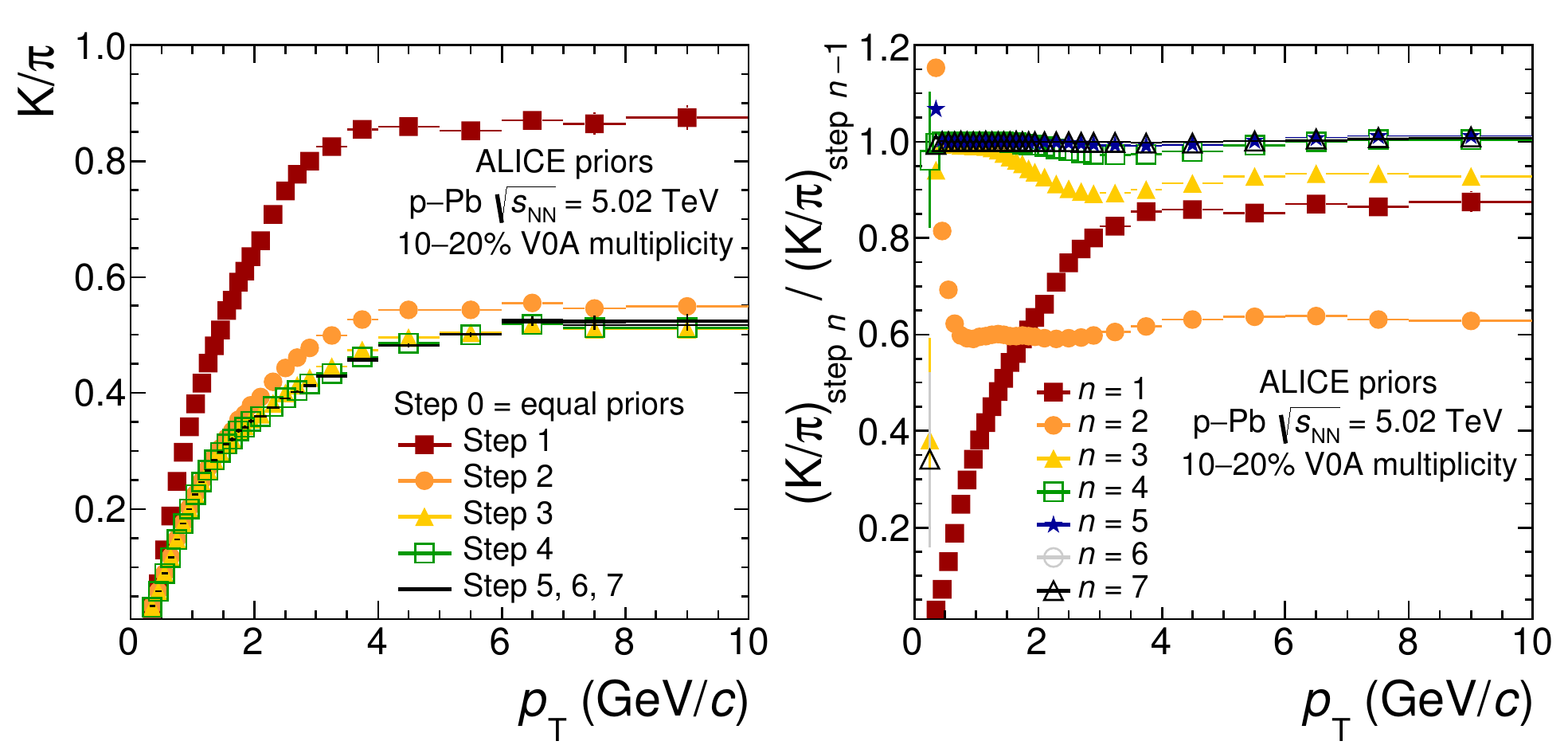}
\caption{An example of the iterative prior extraction procedure for \pPb data (for the 10--20\% V0A multiplicity class). The extracted K$/\pi$ ratio of the priors is shown as a function of \pT at each step of the iteration (left) and as a ratio 
of the value between each successive step (right). Step 0 refers to the initial ratio, which is set to 1.}
\label{fig:priorsConvergence}
\end{figure}

The procedure is then iterated, and the extracted prior values converge progressively with each iteration. The values of the priors are shown in the left panel of~\figref{fig:priorsConvergence} for the K$/\pi$ ratio obtained using data from \pPb collisions. The convergence of this procedure is illustrated
in the right-hand panel, as a ratio of the priors for successive steps. A satisfactory convergence is obtained after 6--7 iterations.
 
A set of priors was obtained for global tracks, defined as tracks reconstructed in both the ITS and TPC. This set is referred to in the following as
$\prior(H_i)_{\rm TPC}$ (the ``standard priors''). This set of priors is then propagated to other detectors 
using ``propagation factors'' $\prop_\alpha$, which are detector-specific and dependent on transverse momentum. In some cases
these multiplicative factors can also be charge-dependent (this is true for EMCal, for example).
$\prop_\alpha$, which is obtained for each detector via Monte Carlo, takes into account 
the particles reaching the outer detectors, as well as the acceptances of the outer detectors and their corresponding energy thresholds.  The abundances measured by TOF and TPC will differ due to these effects. The requirement of a given detector
therefore changes the priors.
For an outer detector $\alpha$, the priors for a track with momentum $\pt$ are determined as
\begin{equation}
\prior(H_i)_{\alpha} (\pt) = \prop_{\alpha}(\pt) \times \prior(H_i)_{\rm TPC}(\pt).
\end{equation}

Priors are currently propagated for TRD, TOF, EMCal and HMPID. Priors were also generated for tracks that are obtained using only ITS hits. These were used for the spectrum analysis at low momentum~\cite{spectra2.76TeV}. In the analyses presented in this paper, the propagation procedure of priors from TPC to TOF was tested extensively.

\begin{figure}[ht]
\centering
\includegraphics[width=0.85\textwidth]{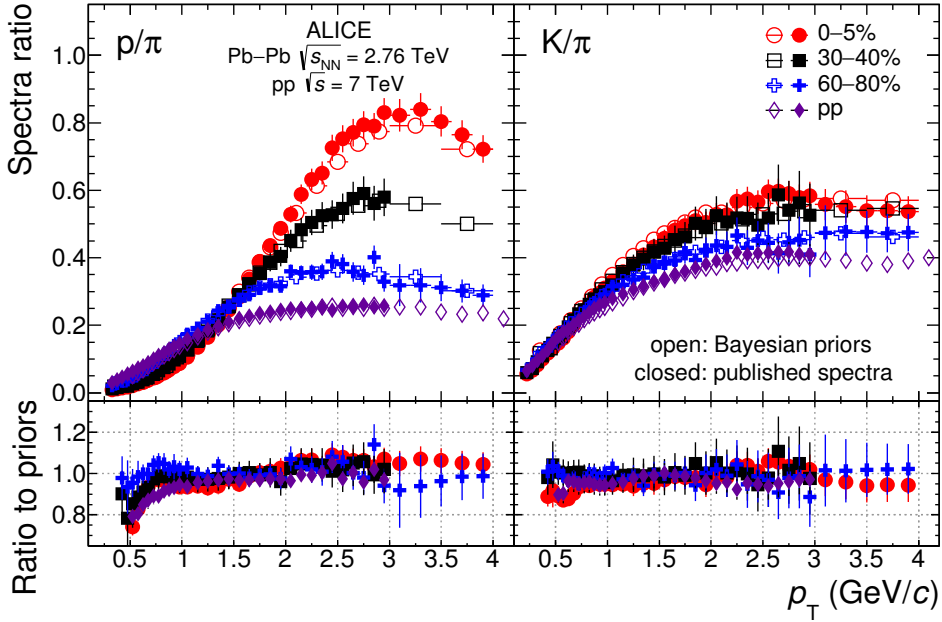}
\caption{The proton/pion ratio (left) and kaon/pion ratio (right), as measured by ALICE~\cite{spectra7TeV,spectraPbPb} using TPC and TOF (filled symbols),
compared with the standard priors as described in the text  (open symbols) for \PbPb and \pp collisions. For \PbPb,  the results are reported for different centrality classes. Particle ratios are calculated for mid-rapidity, $|y|<0.5$. The double ratios (the measured abundances
divided by the Bayesian priors) are shown in the lower panels.}
\label{fig:priorsComparisonPbPb}
\end{figure}

\begin{figure}[t]
\centering
\includegraphics[width=0.85\textwidth]{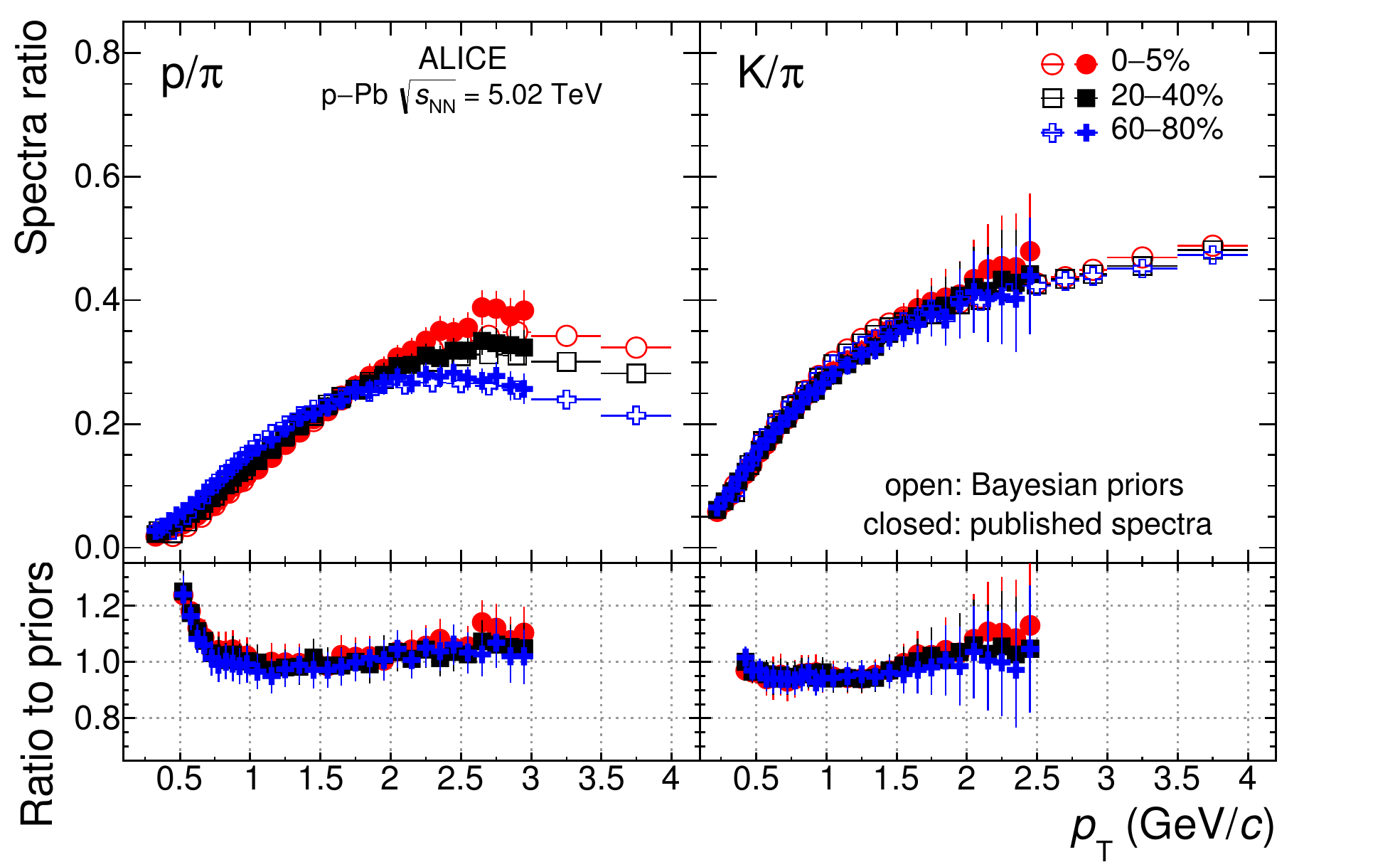}
\caption{The proton/pion ratio (left) and kaon/pion ratio (right), as measured by ALICE ~\cite{spectrapPb} using TPC and TOF (filled symbols),
compared with the standard priors obtained with an iterative procedure (open symbols) for \pPb~collisions for different
V0A multiplicity classes. Particle ratios are calculated for mid-rapidity, $|y|<0.5$ with respect to the centre-of-mass system. The double ratios (the measured abundances
divided by the Bayesian priors) are shown in the lower panels.}
\label{fig:priorsComparisonpPb}
\end{figure}

Since the priors correspond to relative particle abundances, it is possible to directly compare the priors obtained from the iterative procedure with the abundances measured by ALICE~\cite{spectra7TeV,spectraPbPb,spectrapPb}. Comparisons between the priors and the measured p$/\pi$ and  K$/\pi$ ratios are shown in \figref{fig:priorsComparisonPbPb} for \pp and \PbPb collisions at $\sqrts=7$\,TeV and $\sqrtsNN=2.76$\,TeV, respectively, and in \figref{fig:priorsComparisonpPb} for \pPb collisions at $\sqrtsNN=5.02$\,TeV. In some cases the priors cover a larger range than the measurements made by ALICE.

In order to make these comparisons, the priors were corrected by several factors 
to take into account the different selections used in the physics analyses and in the priors computation.
Since the standard priors are provided for the TPC case, only the tracking efficiency correction was applied.
Additional conversions from pseudorapidity intervals (used for the priors) to rapidity intervals (used for the measurement) were also applied, as well as an average feed-down correction derived from the \PbPb~analysis \cite{spectraPbPb}. This feed-down mainly applies to protons from the decays of $\Lambda$ baryons.

The priors (open symbols) and the measured abundances (filled symbols) are consistent with one another within roughly 10\% over a wide momentum range for all centrality ranges (for \PbPb collisions) and V0A multiplicity classes (for \pPb collisions).
The choice to use feed-down corrections evaluated in a given system leads to better agreement at low momenta in \PbPb collisions than in \pPb collisions.
However, at high momenta, where the PID performance is better exploited, the results are independent of this correction. 
The overall level of agreement is satisfactory, as the priors represent a realistic description of the various particle abundances.

It is also important to test whether there is any dependence of the final result on the set of priors used. To perform these tests, it is possible to use alternative sets of priors as well as flat priors, which essentially combine the probabilities from the different detectors without weighting them with Bayesian priors. Slightly different sets of priors can be obtained by varying the track selection parameters. An example of a systematic check on the dependence on varying the set of priors is discussed in \secref{sec:d0res} for the \DZero case.

\subsection{Bayesian PID strategies}\label{sec:strategies}

Once the Bayesian probability for each species (p, K, $\pi$, etc.) has been calculated for a given track, the PID selection may be applied with a variety of selection criteria. The three criteria applied in this paper are:

\begin{itemize}
\item Fixed threshold: The track is accepted as belonging to a species if the probability for this is greater than some pre-defined value. As an example, the choice of a 50\% threshold means that a particle will only be accepted as a kaon if its Bayesian probability of being a kaon is greater than 50\%. Note that this strategy is not necessarily exclusive, as a threshold of less than 50\% could lead to multiple possible identities.
As already discussed, a selection on the Bayesian probability corresponds to a purity requirement of the signal if the priors reflect the
true particle abundances. 
\item Maximum probability: The track is accepted as the most likely species (i.e. the species with the highest probability).
\item Weighted: All tracks reaching the PID step of the analysis are accepted, with a weight different from unity applied to their yield. The weight is defined as the product of the Bayesian probabilities obtained for the tracks involved (e.g. in $\DtoKpi$, 
this is the kaon probability of the negative track multiplied by the pion probability of the positive track). The final result is corrected for the average weight determined in Monte Carlo simulations in the same way as is done for the PID efficiency in other methods.
\end{itemize}

The fixed threshold method is compared with \nsigma PID 
as part of the benchmark analysis (see \secref{sec:V0s}). The maximum probability method is used
in the single-particle spectrum analysis described in \secref{sec:spectra} and for the $\Lc$ baryon analysis in \secref{sec:lcres}. Finally, all three of the aforementioned methods are 
tested for the analysis of $\DtoKpi$ in \secref{sec:d0res}.
\FloatBarrier

\section{Benchmark analysis on two-prong decay channels}\label{sec:V0s}
High-purity samples of identified particles were selected via the study of specific decay channels.
These samples served as a baseline for validating the Monte Carlo tools that are normally used to estimate the
efficiencies and misidentification probabilities
(i.e. the \pid matrix discussed in \secref{sec:pidmethods}) of the Bayesian PID approach.
The methodology developed in this section was also applied to the \nsigma PID approach, providing an
important cross-check. 

\subsection{Description of the method}

The following decays were used to obtain high-purity samples of three different species:
\begin{itemize}
\item $\Kzs \to \piMinus\piPlus$ to study charged pions;
\item $\rmLambda \to \proton\piMinus$ (and respective charge conjugates) to study protons; and
\item $\phimes \to \Kminus\Kplus$ to study charged kaons.
\end{itemize}

The first two cases are true $\Vzero$ decays, comprising two charged prongs originating from a secondary vertex displaced from the interaction point, whereas the daughters of the \phimes meson originate from the primary interaction vertex due to its shorter lifetime. For conciseness, all of these are referred to as $\Vzero$ decays in this paper.

Here we report results based on the \pPb data set, testing the PID method using both TPC and TOF. The \pPb data sample is less affected by background uncertainties than the \PbPb sample when performing fits of the invariant mass spectra, due to the smaller amount of combinatorial background. This is especially true for the \phimes analysis.

Different track selection criteria were applied in order to select daughter particles coming either from the primary vertex (for the \phimes case)
or from a secondary vertex (for the \Kzs and \rmLambda cases).
For a given $\Vzero$, a fit of the combinatorial invariant mass distribution
allows the background to be subtracted and the yield of $\Vzero$ decays to be extracted.
The estimated yield is considered to be a pure sample of a given species (a precise measurement of the total number
of particles of a given species in a given data set).
This estimation was done without applying any PID selections.
Then the exercise was repeated applying
PID selections on each of the two prongs, selecting between pions, kaons and protons.
The comparison with the number of positively identified secondary prongs determines the efficiency and
the misidentification with respect to the values estimated when not applying PID. 

\begin{figure}[htb]
\centering
\includegraphics[width=0.95\textwidth]{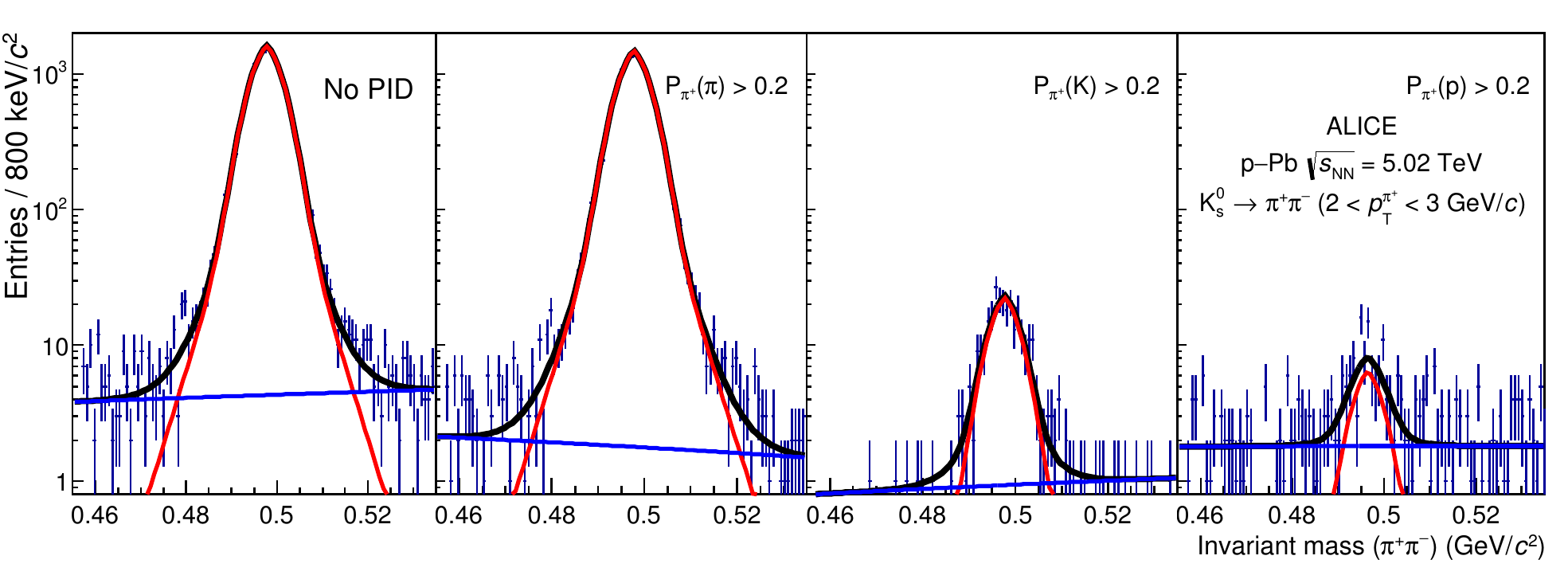}
\caption{$\Vzero$ fits to extract the yield and background for $\Kzs \to \piMinus\piPlus$ in \pPb collisions at $\sqrtsNN=5.02$\,TeV.  From left to right: no PID selections applied
and selecting  pions, kaons and protons using a specific PID strategy (here, Bayesian probability $>$ 0.2). The yield estimated from the second plot from the left (compared with the no-PID yield result) gives a measure of the PID efficiency for the pions, while the remaining ones give information about misidentification.}
\label{fig:piToOther}
\end{figure}

\Figref{fig:piToOther} shows examples of the fitting procedure for the \Kzs invariant masses for $2<\pt^\pi<3\,\GeVc$. From left to right, the panels show the analysis without PID and then requiring the identification of a positive pion, kaon, or proton, respectively.  The \Kzs signal (and background) fitted in the latter three cases, compared with the results without applying PID, allow the extraction of the identification efficiency and misidentification probabilities.

In order to reduce the background for the \phimes~analysis, before starting the procedure, one of the two decay tracks 
was `tagged' using a PID selection requiring compatibility with the kaon hypothesis and then the PID selection under study 
is applied on the other track. The tagging was performed with
a 2$\sigma$ selection on a combination of the TPC and TOF signals. 

The TPC and TOF signals are combined as  $|\mathrm{n}_\sigma^{\rm Comb}(j)| = \sqrt{(\mathrm{n}_\sigma^{\rm TPC}(j)^2+\mathrm{n}_\sigma^{\rm TOF}(j)^2)/2}$.
The same method was also applied on the simulated sample
in order to check the agreement 
between Monte Carlo and data on the estimated quantities. This serves as a validation for the Monte
Carlo estimation of the efficiency of a given PID strategy, as well as for the subsequent corrections required in order to extract physics results.

\subsection{Comparison of PID efficiencies between data and Monte Carlo}

\begin{figure}[htb]
\centering
\includegraphics[width=0.95\textwidth]{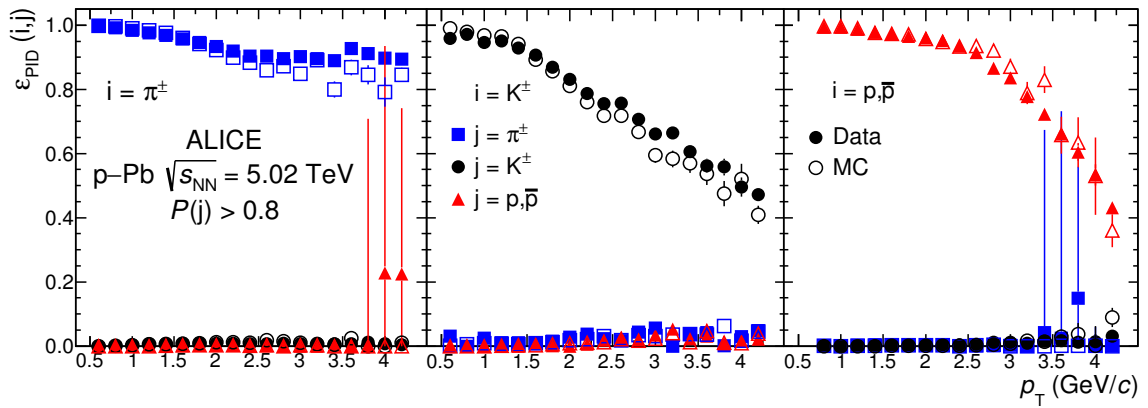}
\caption{\pid matrix elements in \pPb collisions after selection with a Bayesian probability greater than 80\%. 
Comparisons with Monte Carlo (open symbols) are also shown.}
\label{fig:pid80}
\end{figure}

The PID matrix elements obtained for different Bayesian probability thresholds are presented here.
An example for the highest purity case considered in this work (Bayesian probability greater than 80\%) is shown in \figref{fig:pid80}.
Each plot represents a row of the \pid matrix for a given species $i$,
with the $i=j$ points corresponding to the PID efficiencies ($\epsilon_{ii}$) and the $i\neq j$ points corresponding to the misidentification probabilities ($\epsilon_{ij}$).
The matrix was evaluated separately for positively and negatively charged tracks. As no difference was found between the two cases, the results shown here were averaged over both charges.
For Monte Carlo, the PID hypothesis was tested both via the true particle identity available in the simulation
and by applying the same procedure as for the data.

\begin{figure}[htb]
\centering
\includegraphics[width=0.95\textwidth]{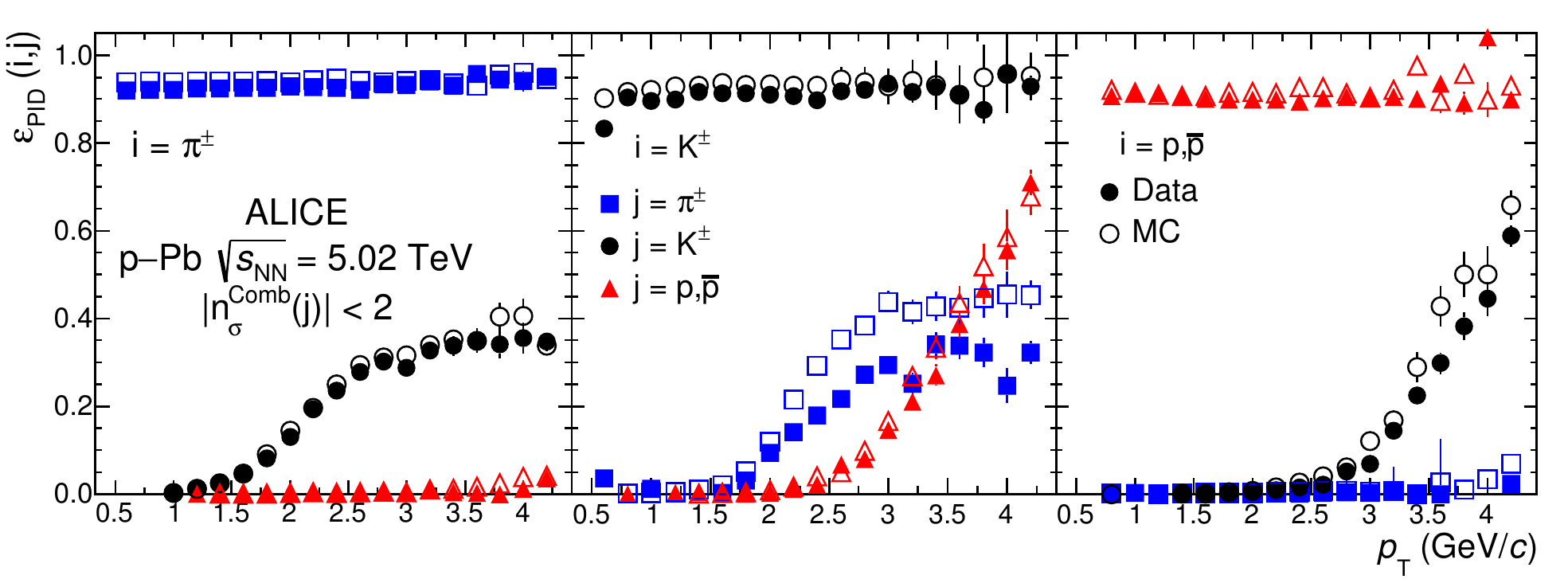}
\caption{\pid matrix elements in \pPb collisions after selection with a 2$\sigma$ selection on the combined TPC and TOF signal. 
Comparisons with Monte Carlo (open symbols) are also shown.}
\label{fig:2sigmaCut}
\end{figure}

As can be seen from \figref{fig:pid80}, the efficiencies
and misidentification probabilities can be evaluated very precisely.
The agreement between data and Monte Carlo is good, both in shape and absolute value. The general features
of the PID strategies are also described well, and behave as expected: using a high threshold maximises the purity,
but also sharply reduces the efficiency. Nevertheless, even for $\prob\ge80\%$, where the efficiency estimate is more sensitive to the description of the detector responses in the simulations,
the agreement with Monte Carlo remains within 5\% below 3\,\gevc.

The analysis was repeated using 2$\sigma$ and 3$\sigma$ selections on the combined TOF and TPC signals, as discussed above, to identify the three hadron
species. 
The result is shown in \figref{fig:2sigmaCut} for the 2$\sigma$ case.  As expected, the \pid matrix elements have different values from the 80\% Bayesian probability threshold (given that the $\rm n_\sigma^{Comb}$ selection is somewhat more inclusive),
and the probability of misidentification increases accordingly. The agreement between Monte Carlo and data for
the misidentification probabilities is worse in the \nsigma case than in the Bayesian case for kaons misidentified as pions. 
However, there remains a good agreement between Monte Carlo and data overall. The efficiencies are below the values expected from a perfectly
Gaussian signal; a Monte Carlo or data-driven evaluation of the PID strategy efficiency is therefore
mandatory. The non-Gaussian tail of the TOF signal~\cite{TOFperf}, caused by charge induction on pairs of neighbouring readout pads, plays a significant role in this discrepancy. The mismatch fraction is also not 
negligible in \pPb collisions, being $\sim$2\% above 1\,\gevc. 

\begin{figure}
\centering
\includegraphics[width=0.95\textwidth]{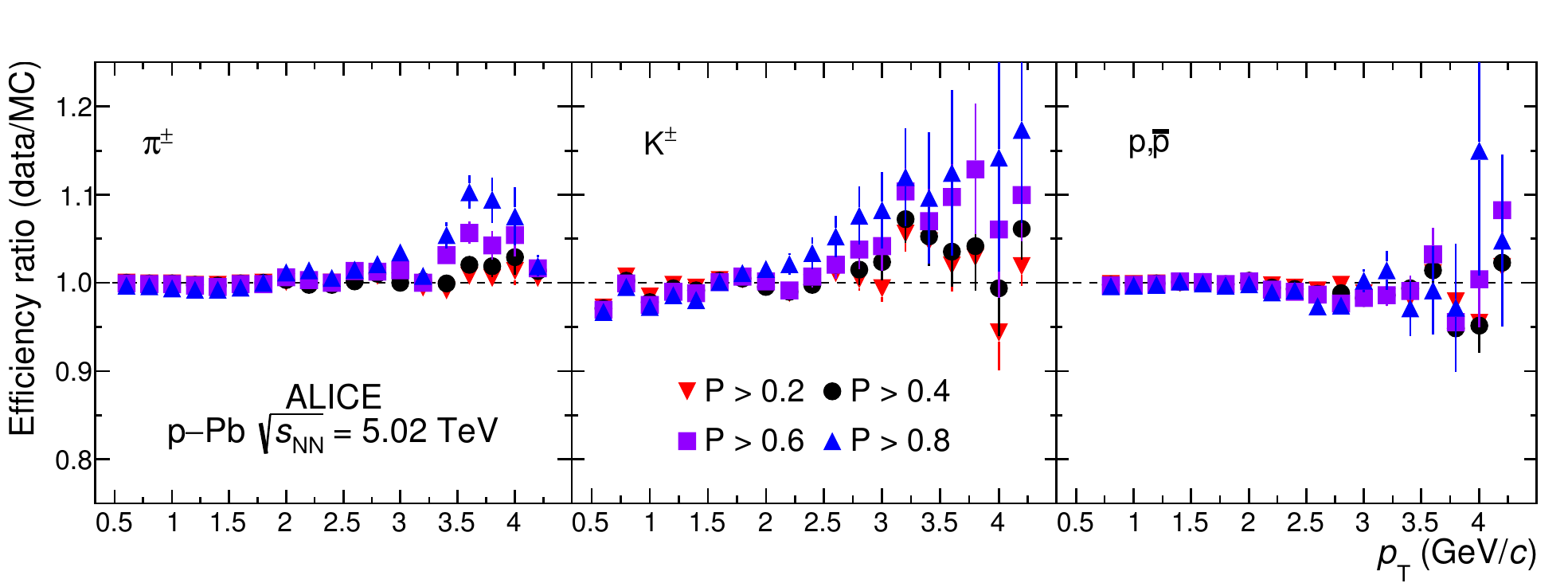}
\caption{Data/Monte Carlo ratios of PID efficiencies for pions, kaons and protons in \pPb collisions, extracted using different Bayesian probability thresholds.}
\label{fig:effBayesianSummary}
\end{figure}

\begin{figure}
\centering
\includegraphics[width=0.95\textwidth]{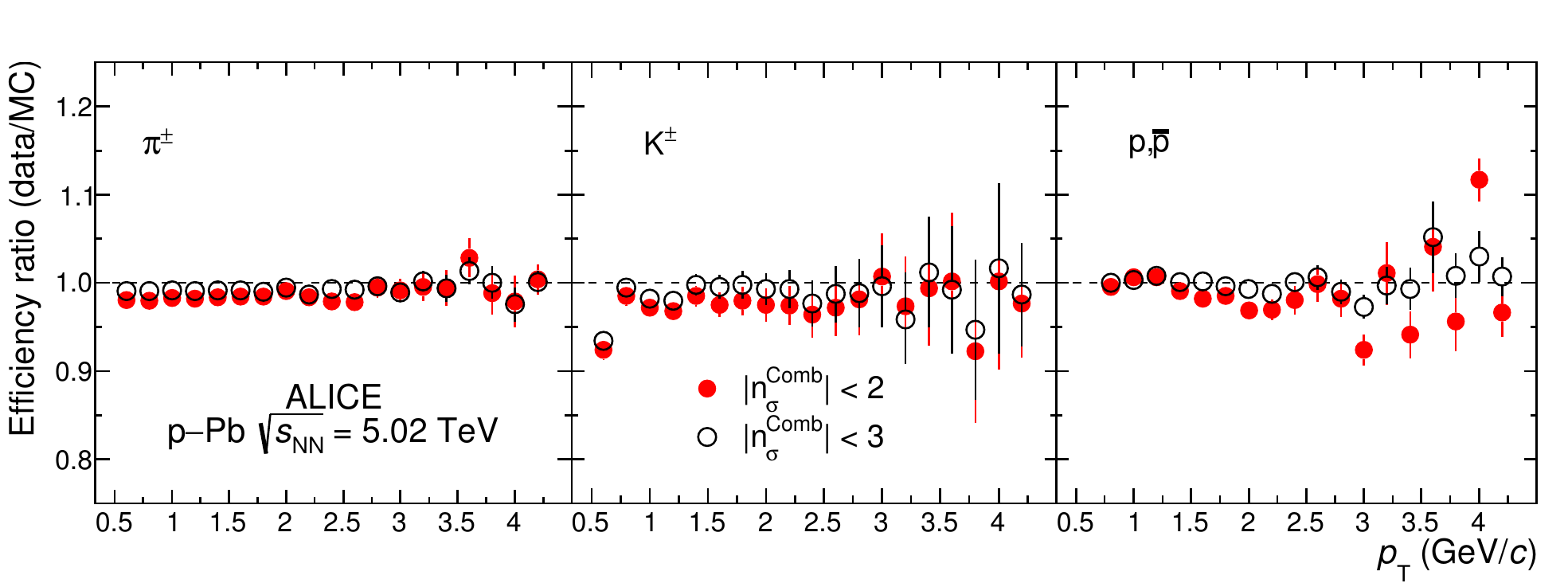}
\caption{Data/Monte Carlo ratios of PID efficiencies for pions, kaons and protons in \pPb collisions, extracted using 2- and 3$\sigma$ selections on the combined TPC and TOF signal.}
\label{fig:effGaussianSummary}
\end{figure}

For a more detailed comparison, the ratios
between data and Monte Carlo are presented in Figs.~\ref{fig:effBayesianSummary} and~\ref{fig:effGaussianSummary} for efficiencies obtained using different Bayesian probability thresholds,
and a 2$\sigma$ or 3$\sigma$ PID selection, respectively. The agreement is very similar (within $\sim$5\%) for both methods. A 
larger difference is seen for kaons when using a very high Bayesian probability threshold (corresponding to
a ``high purity'' strategy). However, such an approach could still be beneficial in physics analyses
where an efficiency correction is not needed (such as analyses investigating Bose-Einstein correlations~\cite{femto} or the flow of identified particles~\cite{flow}).
The uncertainties reported in these plots are purely statistical.
Apart from the case of kaons extracted from \phimes decays, the statistical uncertainties on the Monte Carlo simulations give the largest contribution to the uncertainties shown here.
The \phimes-meson invariant mass plots are affected by a larger combinatorial background in the data sample when PID is not requested (i.e. in the denominator of the efficiency).
As the ratios of data to Monte Carlo for both Bayesian and \nsigma PID are close to unity in~\figrefs{fig:effBayesianSummary} and~\ref{fig:effGaussianSummary},
it can be concluded that the systematic uncertainties from the PID procedure are well under control in both cases.

In summary, the \Vzero analysis technique described in this section serves not only as a validation of the quality of
the Monte Carlo description of the various detector responses, but can also be used in data and simulations to validate different PID strategies and track selections
for any kind of analysis. Finally, it provides a tool to investigate the systematic uncertainties that arise due to the PID selection.

\FloatBarrier

\section{Bayesian approach applied to physics analyses}
\label{results}

In this section we present validations of the Bayesian PID approach for two analyses already published by the ALICE Collaboration in pp collisions at $\sqrts=7$\,TeV: identified pion, kaon and proton spectra~\cite{spectra7TeV} and \DtoKpi~\cite{alice2011charmDmesonsInPP7TeV}. While the previous papers remain the proper references for the extraction of the 
physical quantities such as cross sections, and for theory comparisons, this paper shows the results obtained by 
applying the Bayesian approach to the PID part of those analyses.

\subsection{Identified hadron spectra}
\label{sec:spectra}

The consistency of the Bayesian PID technique was tested using the tools
described in the previous sections to obtain the \pt spectra of pions, kaons and protons. This analysis used 
a data sample of $1.2 \times 10^{8}$ inelastic \pp collisions at $\sqrt{s} = 7$\,\TeV
that was collected in 2010.

The results are compared here with similar measurements that were already reported in~\cite{spectra7TeV}.
In the quoted paper, different PID techniques were used depending on the
detectors involved in different \pT ranges (\nsigma for ITS only and for the combined TPC
and TOF signals, and unfolding techniques for TOF and HMPID separately). Charged kaon spectra
were also measured via the identification of their decays (measurement of the kink topology).  Full details of the original analysis, including the event selection criteria (which were also used here) and the procedure used to merge the various PID techniques and detectors,
can be found in~\cite{spectra7TeV}.

The efficiency, detector acceptance, and other correction factors, were estimated 
using Monte Carlo samples that simulated the detector conditions run-by-run. The simulation was based on the PYTHIA6.4 event generator~\cite{pythia} using the Perugia0 tune~\cite{perugia}, with events propagated through the detector 
using GEANT3~\cite{geant3}.

Only the TPC and TOF detectors were used for PID.
The particle species were identified using the maximum probability method outlined in \secref{sec:strategies}.
The identification of charged hadrons used different detector combinations in different momentum ranges, as shown in Table~\ref{ta:spectraCut}. Although a Bayesian approach does not necessarily require such a 
division, it was chosen in order to make a closer comparison with the analysis presented in~\cite{spectra7TeV}. In particular, the TOF
efficiency drops very steeply at low momentum due to the acceptance, meaning that the systematic uncertainty on the TOF efficiency would become dominant and
make the comparison difficult.

\begin{table}[h]
\begin{center}
\begin{tabular}{ c c c }
\hline 
Hadron & TPC & TPC--TOF \\ 
$\pi$  & 0.2 $\leq\pt\leq$ 0.5 & 0.5 $\leq\pt\leq$ 2.5 \\ 
K & 0.3 $\leq\pt\leq$ 0.45 & 0.45 $\leq\pt\leq$ 2.5 \\ 
p  & 0.5 $\leq\pt\leq$ 0.8 & 0.8 $\leq\pt\leq$ 2.5 \\ \hline
\end{tabular}
\caption{PID detectors and transverse momentum ranges (in \gevc) used for the analysis of identified hadron spectra.}
\label{ta:spectraCut}
\end{center}
\end{table}

The PID efficiency is higher than 95\% for pions and protons, while for kaons it begins to decrease from 100\%
at 1\,\gevc to 75\% at 2.5\,\gevc. 
The misidentification percentage is below 5\% for pions and protons for all \pt, and reaches 20\% for kaons at $\pt = 2.5\,\GeVc$.
In order to avoid the dependence of the corrections on the relative abundances of each hadron species in the event generator, the spectra were corrected for their respective PID 
efficiencies and for contamination using the \pid matrix method described in \secref{sec:effcontam}.
A $4\times4$ matrix (also including electrons) was defined for each \pT interval.
These matrices were then inverted and used in Eq.~\ref{eq:trueabund}
in order to obtain the spectra.
In addition, the spectra were corrected for the tracking, TPC--TOF matching and primary vertex determination efficiencies.
The contributions from secondary particles that were not removed by a selection based on the distance of closest approach to the vertex were determined using a data-driven method, as explained in~\cite{spectra7TeV}, 
and were subtracted from the final spectra.   

\begin{figure}[th]
\begin{center}
\includegraphics[width=0.95\textwidth]{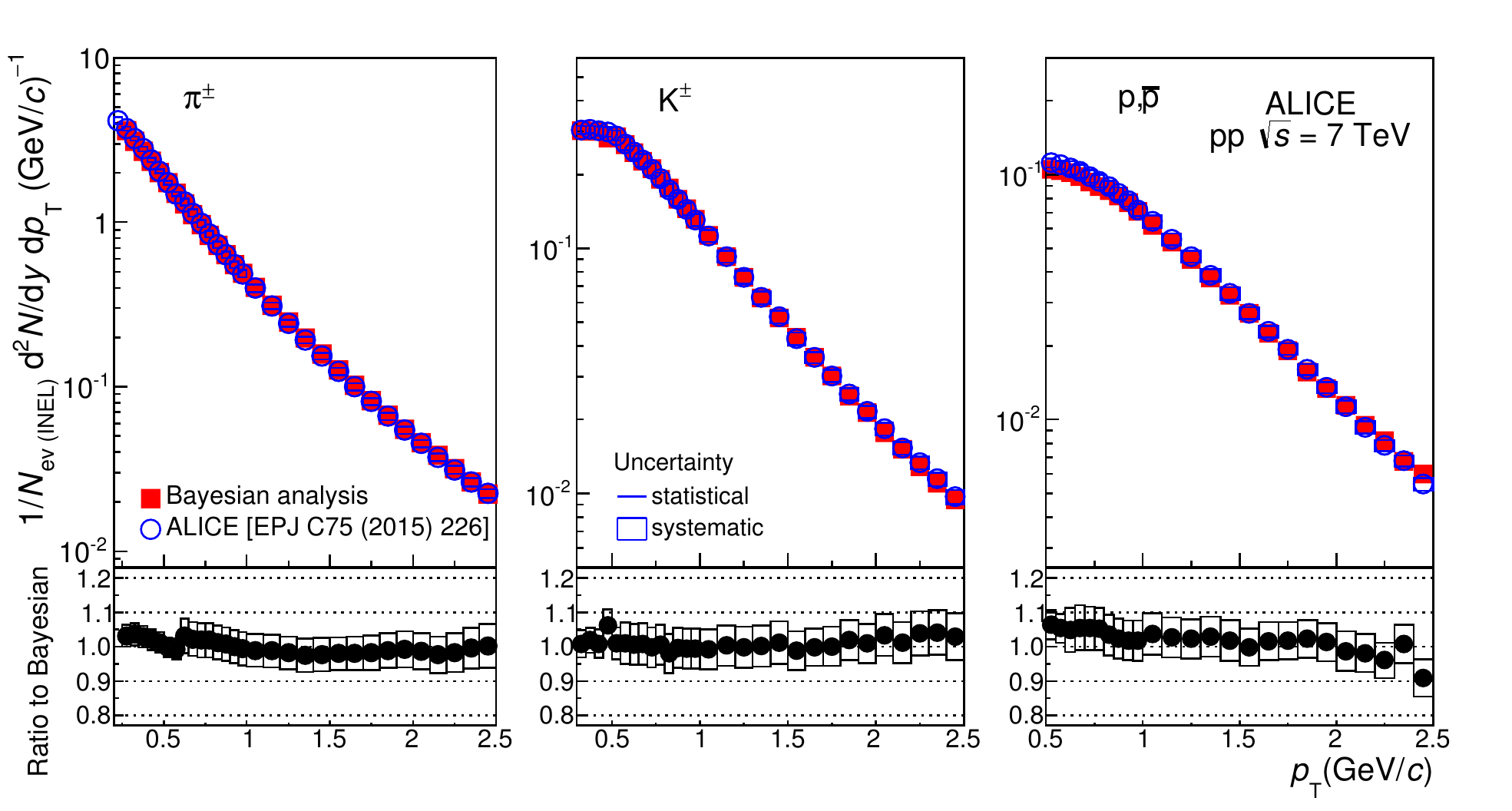}
\end{center} 
\caption{Identified particle spectra from the Bayesian analysis, compared with the measurement reported by ALICE in pp collisions at 7\,\TeV~\cite{spectra7TeV}.}
\label{fig:compareSpectra}
\end{figure}

\Figref{fig:compareSpectra} compares the minimum-bias charged hadron spectra from \pp collisions at $\sqrt{s}=7$\,\TeV obtained from this analysis with the 
published result. A very good agreement within the uncertainties can be observed between the \pT~spectra obtained using these different approaches. For this comparison, the statistical uncertainties are shown for both analyses, while the systematic uncertainties that are not only related to PID 
were only considered for the results published in~\cite{spectra7TeV}. The ratios of the spectra, presented in the lower panels of \figref{fig:compareSpectra}, show an agreement within $\pm5\%$ for all species.

A further check was performed testing the stability of the method against the priors used. The analysis was repeated using flat priors, i.e. equal probabilities for the four particle species that are included in the PID matrix (electrons, pions, kaons and protons). The muon priors were set to zero, their percentage being negligible (with standard priors, they are estimated to be less than 2\% with respect to all other particle species). Despite considering quite an extreme case here in terms of varying the priors, the pion and proton abundances were consistent with the result obtained with standard priors within 3\%. A decrease in the estimated pion abundance at $\pT>1\,\gevc$ resulted in a variation of up to 10\% for kaons in some \pT intervals. The observed variations can be interpreted as being due to remaining uncertainties in the detector response; a naive set of priors, such as a set of flat priors, is generally expected to amplify such effects.

\FloatBarrier

\subsection{Analysis of $\DtoKpi$}
\label{sec:d0res}

This section presents a comprehensive overview of a variety of Bayesian PID strategies applied to the analysis of \DtoKpi (and charge conjugates) in pp collisions at $\sqrt{s}=7$\,\tev.
 The analysis was based on a data sample of roughly $3\times10^8$ events collected during \run{1}.
The geometrical selections on the displaced decay vertex topology matched those used in the ALICE measurement of $\DtoKpi$ in pp collisions reported in\,\cite{alice2011charmDmesonsInPP7TeV}.

In order to make a detailed assessment of the Bayesian method, this analysis compares the results obtained using each of the PID strategies outlined in \secref{sec:strategies}. In each case, the PID method relied on selecting an oppositely charged pair of tracks corresponding to a kaon and a pion.
The TPC and TOF detectors were used in conjunction with one another. For tracks without TOF information, the PID was based on information from the TPC only. 

For the fixed-threshold method, probability thresholds of 40\%, 50\%, 70\%, and 80\% were tested. Note that in the case of the 40\% threshold, there is the possibility that a daughter track may be compatible with both the kaon and pion hypothesis; in such cases, the track was accepted as possibly belonging to either species.
For the maximum-probability and fixed-threshold methods, once the daughter tracks were analysed, the candidate was accepted or rejected according to the following criteria:
\begin{itemize}
\item  if both daughters were identified as possible kaons, the candidate was accepted both as a \Dzero and a \Dzerobar;
\item if one daughter was identified as a kaon and the other as a pion, the candidate was accepted as a \Dzero if the negative track was a kaon and the positive track was a pion, and vice-versa for \Dzerobar;
\item if neither daughter was identified as a kaon, the candidate was rejected;
\item if either daughter was not compatible with the kaon or the pion hypothesis, the candidate was rejected.
\end{itemize}

The weighted method was implemented as defined in \secref{sec:strategies}, whereby the invariant mass distributions were filled for each candidate with weights $W_i$. These weights are defined as

\begin{equation}\begin{aligned}   W_{\Dzero} = P_{\Kminus}& \times P_\piPlus,\text{ and}\\ 
 W_{\Dzerobar} = P_{\Kplus}& \times P_{\piMinus},
\end{aligned}\end{equation}
where $P_i$ corresponds to the Bayesian probability assigned to each track for a given species $i$.
In this case, the PID efficiency from Monte Carlo corresponded to the average weight that was assigned to simulated \Dzero mesons in each $\pt$  interval.

Once the invariant mass distributions in each $\pt$ interval had been obtained, the yields were fitted using a Gaussian function for the signal and an exponential function to estimate the background. The raw signal was extracted according to the integral of the fit to the signal distribution.
In order to ensure the stability of the fitting procedure, the width and mean of the Gaussian functions were kept the same in each \pt interval for all PID methods. To do this, the fit parameters were first determined for the \nsigma PID method in each $\pt$ interval, and then fixed to those values in the fitting procedure for each other PID method.

\begin{figure}[htb]
\centering
\includegraphics[width=0.95\textwidth]{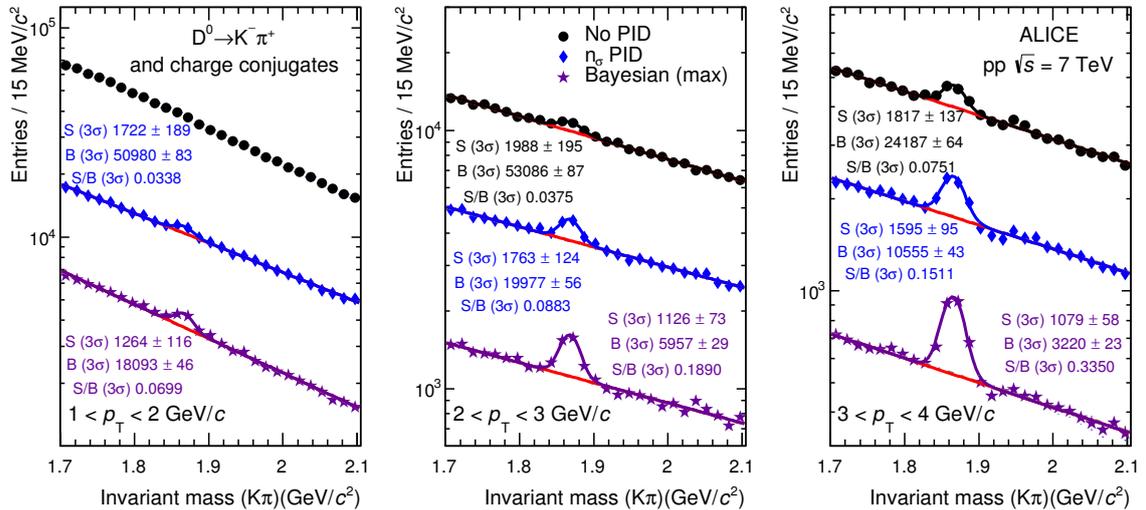}
\caption{A comparison of the invariant mass distributions in three $\pt$ intervals for \Dzero candidates obtained without PID, with \nsigma PID, and with Bayesian PID using the maximum probability condition. Due to the low statistical significance, it was not possible to extract a stable signal without PID for $1<\pt<2\,\GeVc$, therefore this fit and its results are not shown.}
\label{fig:d0invmasscomp}
\end{figure}

Example invariant mass distributions obtained without PID, with \nsigma PID, and with Bayesian PID using the maximum-probability strategy are shown in \figref{fig:d0invmasscomp}, with a logarithmic scale on the vertical axis. 
There is a clearly visible increase in the signal-to-background ratio when using \nsigma PID as compared to the analysis without PID, and a further increase when applying a Bayesian selection with the maximum-probability strategy. Due to the large background and low statistical significance, it was not possible to extract a stable yield for $1<\pt<2$~\GeVc without PID. For this reason, the results obtained without PID are excluded for this \pt interval in the following comparisons.

\begin{figure}[htb]
\centering
\includegraphics[width=0.49\textwidth]{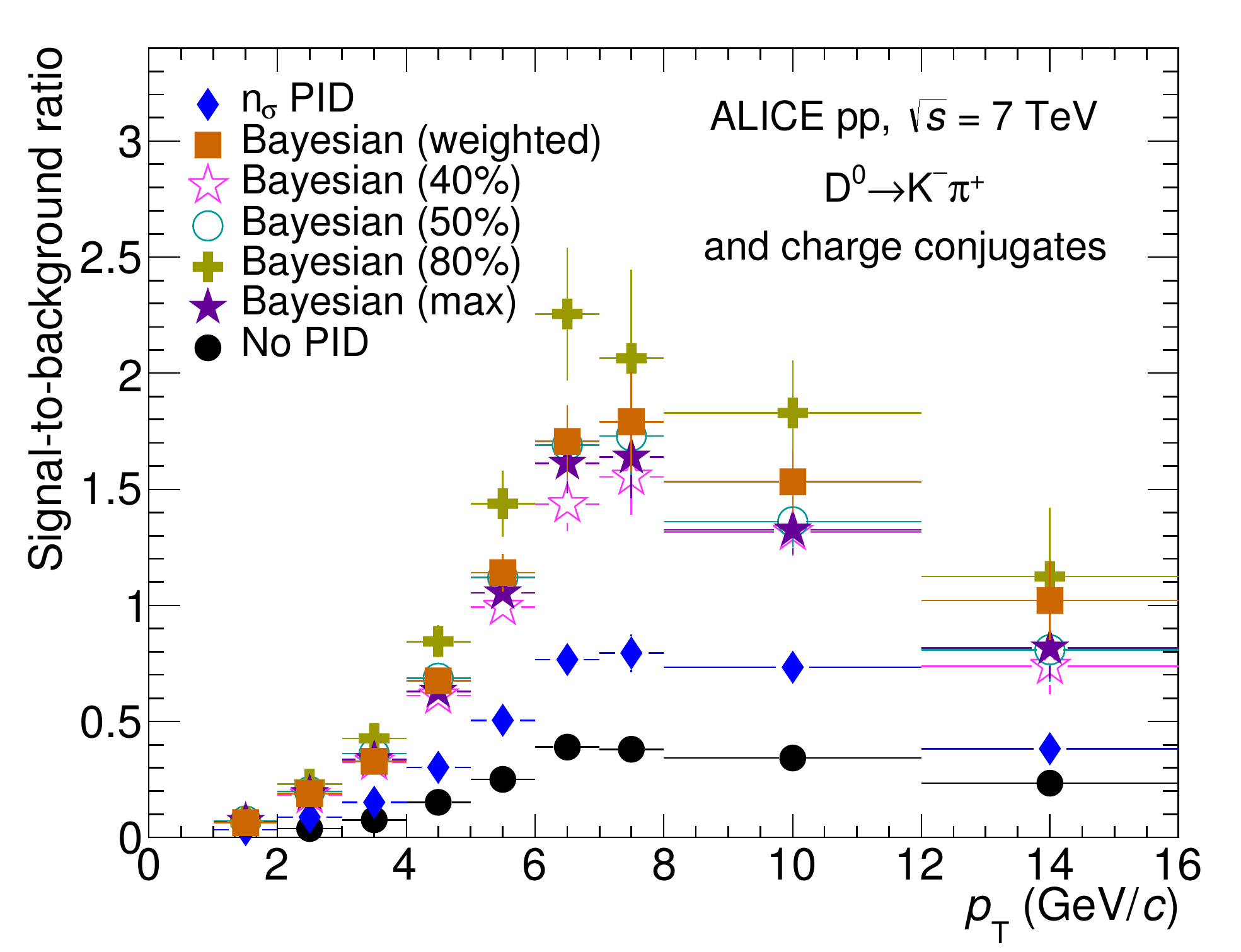}
\includegraphics[width=0.49\textwidth]{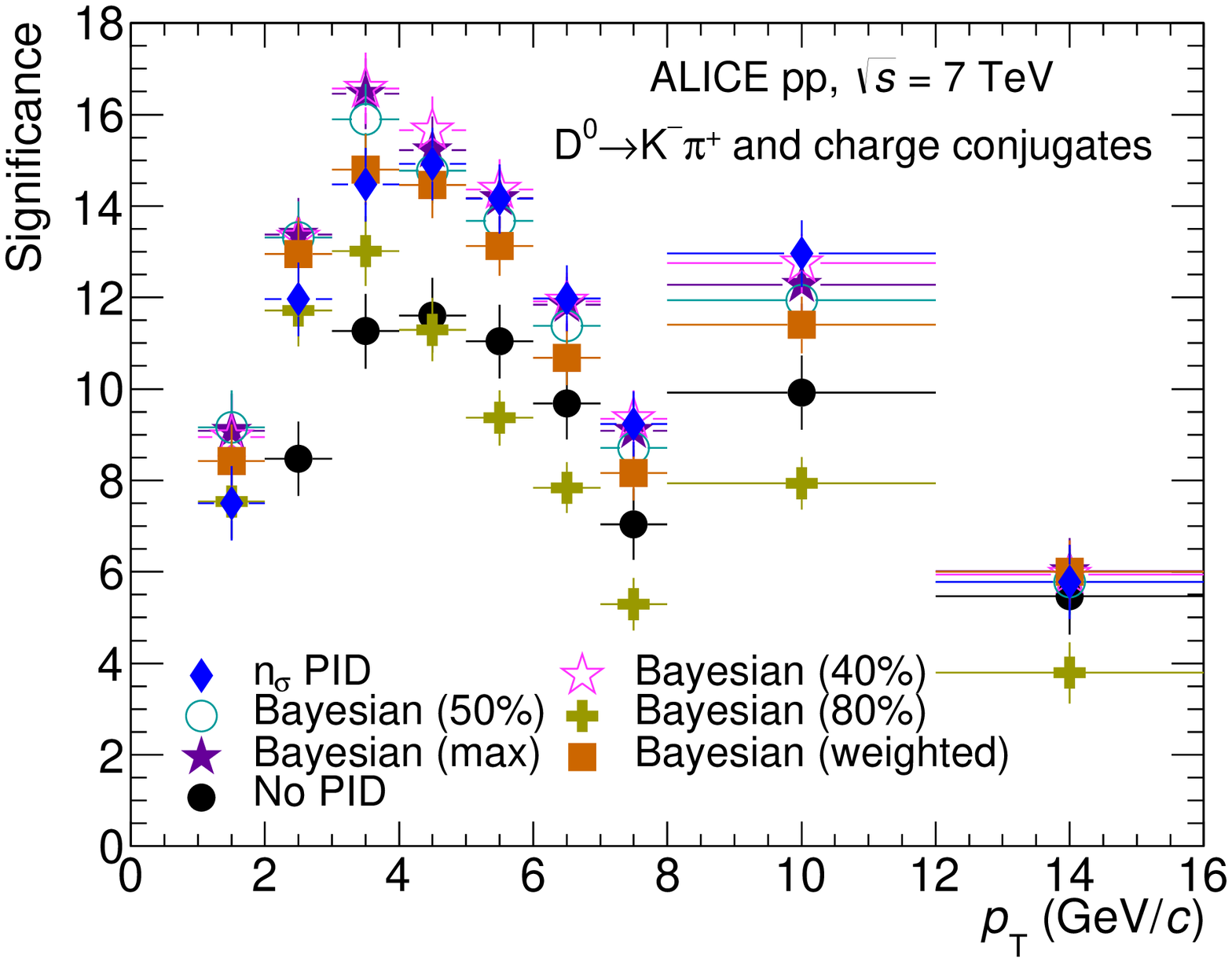}
\caption{(Left) Signal-to-background ratio and (right) statistical significance as a function of $\pt$ for various methods of particle identification. Note that the increase in significance at $8 < \pt < 12\,\gevc$ is an effect of the width of the $\pt$ interval increasing from 1 to 4\,\gevc.}
\label{fig:d0_sigbacksignif}
\end{figure}

Full comparisons of the signal-to-background ratio and significance obtained for different PID strategies are shown in \figref{fig:d0_sigbacksignif}. The statistical significance is defined as the signal divided by the square root of the sum of the signal and background.
The signal-to-background ratio increases significantly for all \pt when using the \nsigma PID approach, as compared to not making a PID selection.
A similar increase, of a factor of at least 2, was seen in each of the Bayesian methods when compared with \nsigma PID.

As expected, the statistical significance of the signal was higher for the \nsigma PID than when no PID selection was applied.
At low $\pt$, all but the strictest Bayesian selections (i.e. 80\% probability threshold) gave a further increase in significance over the \nsigma PID method, while at higher $\pt$ the significance of these methods became similar to that of \nsigma PID.
The 80\% probability threshold yielded a lower statistical significance than \nsigma PID for $\pt > 3\,\GeVc$, whereas the performance of the two methods is equivalent for $\pt\leq3\,\GeVc$.

\begin{figure}[htb]
\centering
\includegraphics[width=0.49\textwidth]{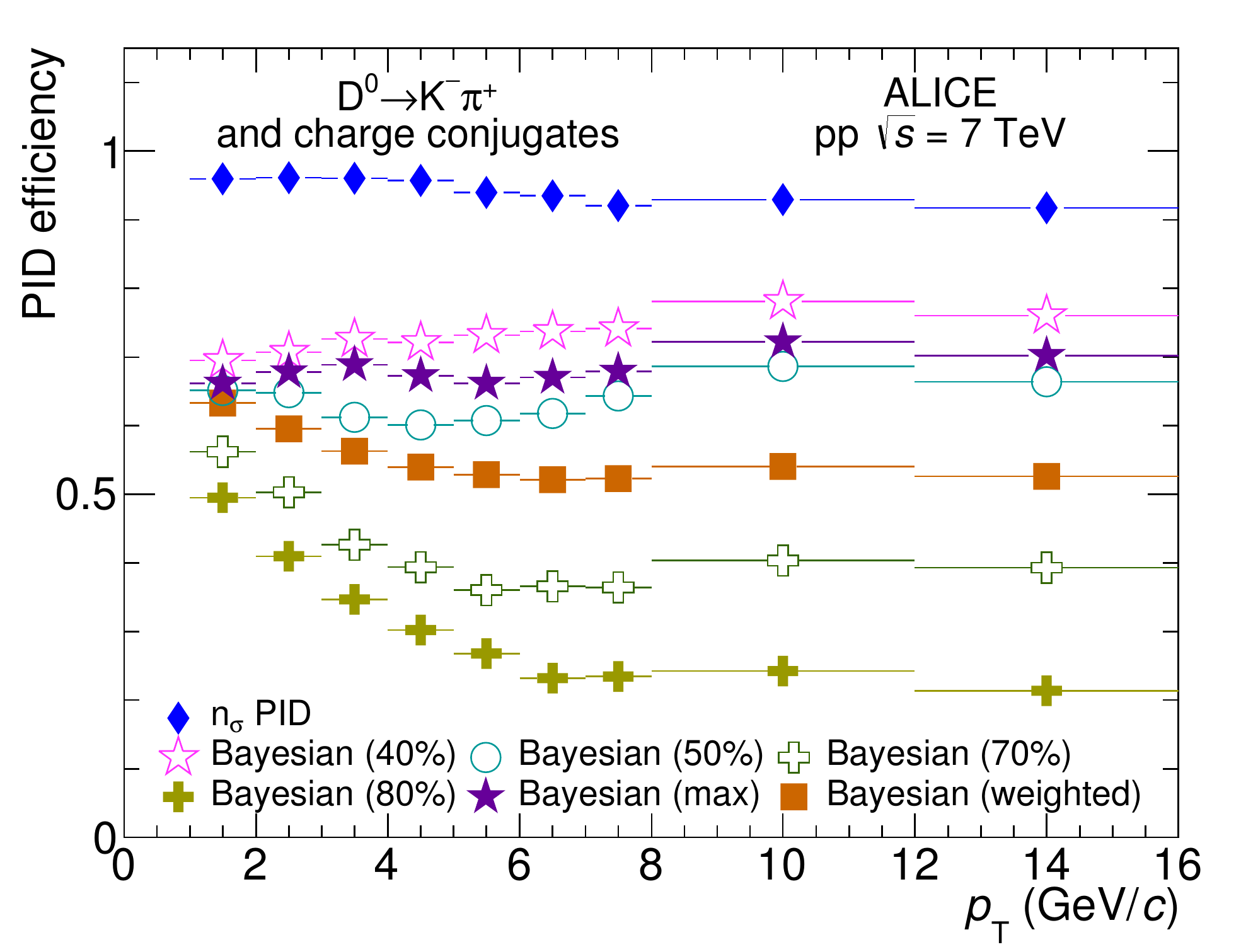}
\caption{A comparison of the PID efficiencies for $\DtoKpi$ obtained using various PID strategies, as a function of $\pt$.}
\label{fig:d0_effs}
\end{figure}

The efficiencies for each particle identification method were obtained using charm-enriched Monte Carlo simulations, as described in~\cite{alice2011charmDmesonsInPP7TeV}. 
The PID efficiency was defined as the proportion of simulated D mesons that passed the PID selection criteria, having already passed the other selections.
In the case of the weighted Bayesian PID method, as previously mentioned, the PID efficiency corresponded to an average weight for true \Dzero candidates, and for the purposes of corrections was used in the same way as the other PID efficiencies. The PID efficiencies obtained for each method are shown in \figref{fig:d0_effs}.

When a stricter PID selection was applied (for example, an 80\% probability threshold as compared to a 70\% threshold), the PID efficiency fell correspondingly.
All of the Bayesian methods gave a lower PID efficiency than the \nsigma method. The observed significances and efficiencies highlight a benefit and a potential drawback in using a strict Bayesian PID strategy: while a tighter PID selection gives a yield of higher purity, it also requires a greater reliance on the Monte Carlo simulation to give an accurate correction due to the reduced PID efficiency. However, it is important to note that this
is not necessarily inherent to the Bayesian PID method. For example, a very narrow \nsigma selection (e.g. at 0.5$\sigma$) would also yield a lower efficiency, leading to a similar degree of reliance on the Monte Carlo simulations; even a 2$\sigma$ selection reduces the PID efficiency to 80\%.

\begin{figure}[htb]
\centering
\includegraphics[width=0.49\textwidth]{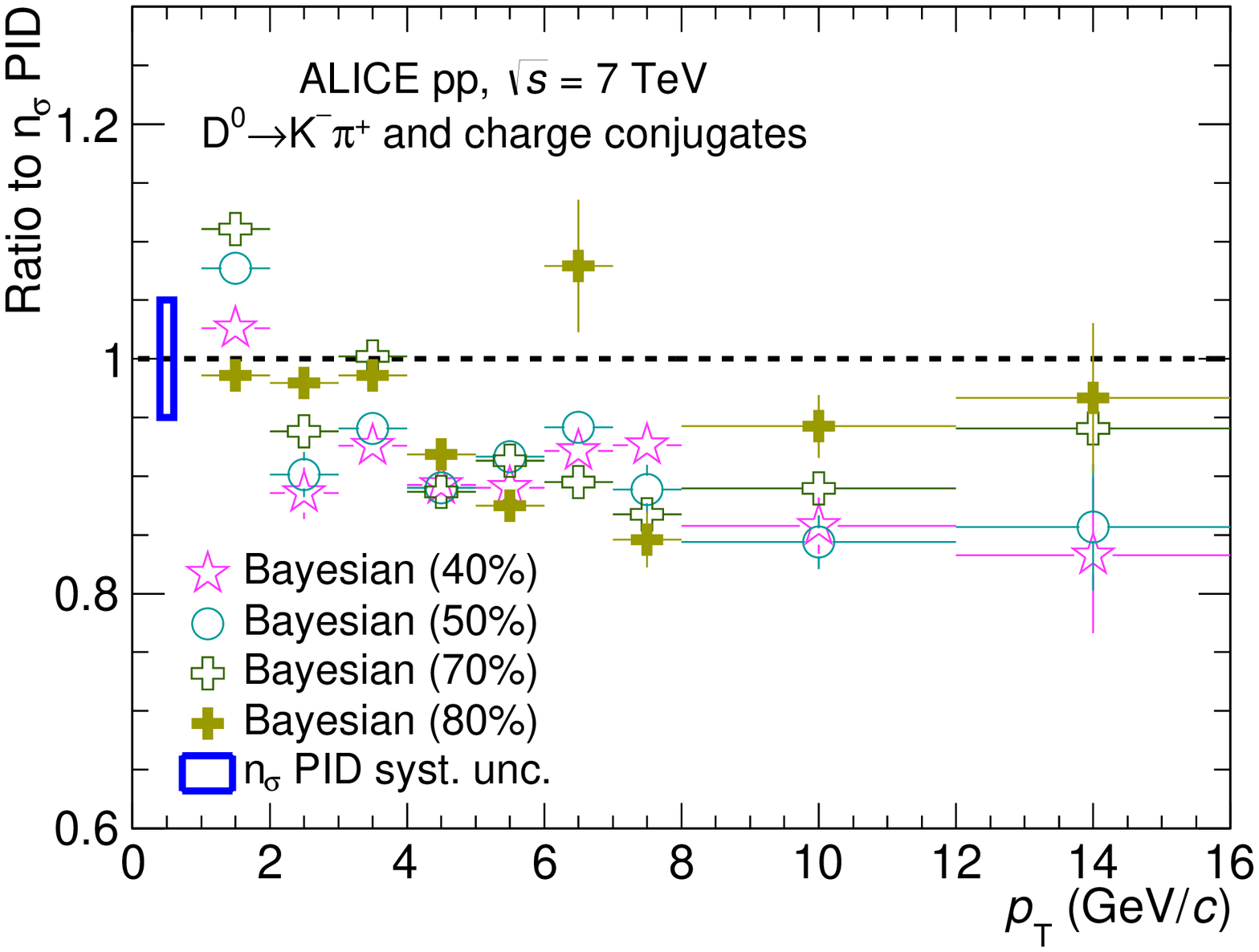}
\includegraphics[width=0.49\textwidth]{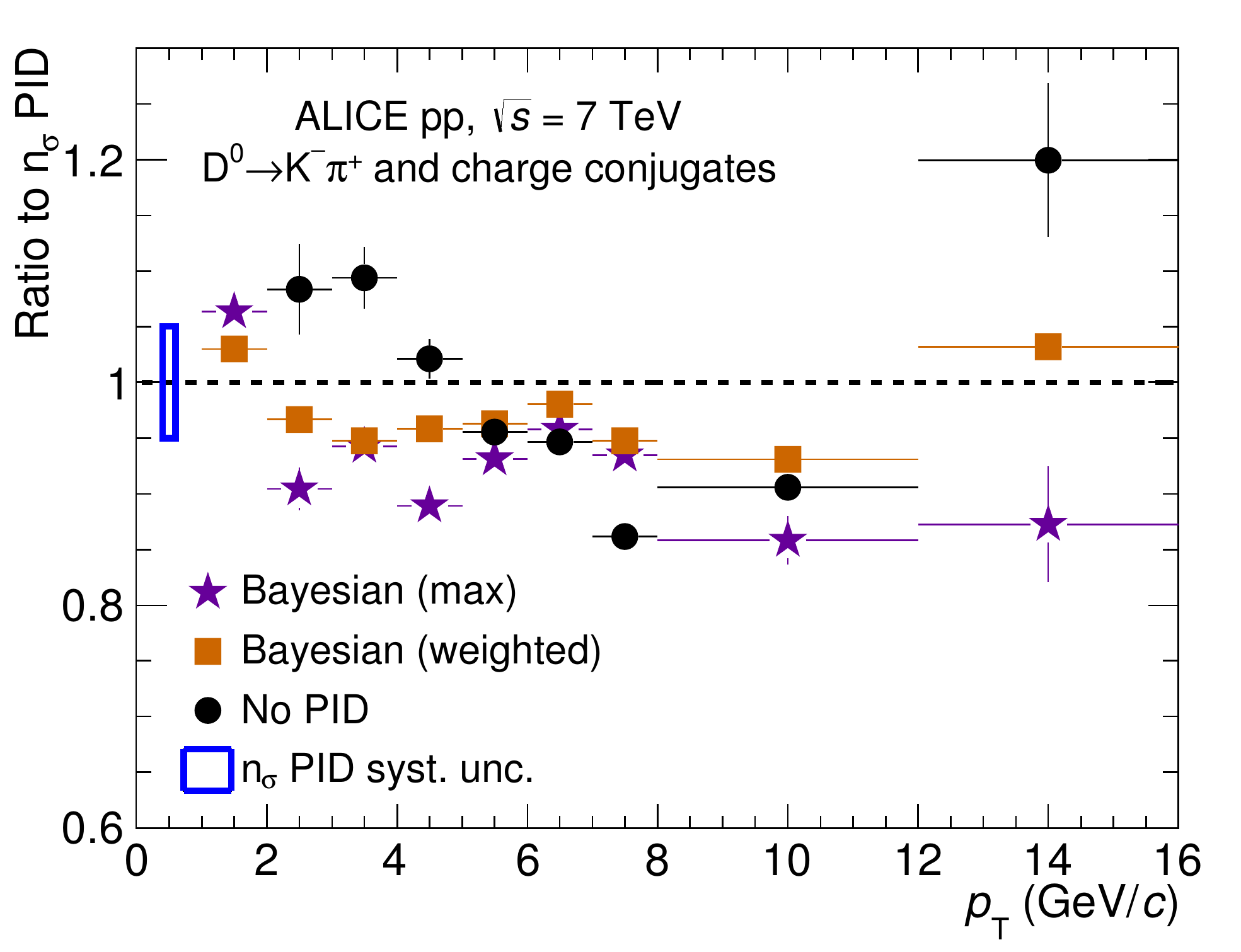}
\caption{Ratios of corrected yields obtained using various Bayesian PID methods to that obtained using \nsigma PID, for (left) fixed probability thresholds, and (right) maximum probability and weighted Bayesian PID, and no PID. The 5\% systematic uncertainty on the \nsigma PID method is shown as a blue box at $0.5\,\GeVc$.}\label{fig:d0rescomp}
\end{figure}

The corrected yields obtained using each particle identification method are compared in \figref{fig:d0rescomp}. In each case, the ratio shown is the corrected yield divided by that obtained using \nsigma PID. As the candidates selected by each of the Bayesian methods are a subset of those selected by the \nsigma approach, the uncertainties on the corrected yields are assumed to be fully correlated and so almost cancel out in the ratio.

It was found that there is a relatively good agreement between the Bayesian PID method and the \nsigma PID method, even when very different probability thresholds are applied. Averaging over \pt, the fixed probability threshold methods give a systematic difference between 5 and 9\%, increasing for the lower probability thresholds.
For the weighted and maximum-probability methods, the average deviations from \nsigma are 3\% and 7\%, respectively. In every case this difference is of a similar order to the systematic uncertainty of the \nsigma PID approach (${\sim}5\%$)~\cite{alice2011charmDmesonsInPP7TeV}.

In effect, the difference of the corrected yield from the \nsigma result represents the systematic uncertainty due to the choice of Bayesian PID method. Therefore it can be concluded that the PID systematic uncertainties for the Bayesian approach are not much larger than those from the \nsigma approach. The observed deviations also contain systematic contributions related to the yield extraction procedure.

A final check was made to ensure that the results remained independent of the choice of the prior probability distribution used. Ideally, the requirement of an external input should guide the measurement, and fluctuations in these priors should not have a substantial effect on the final result that is obtained. 
In order to test this, two modified sets of priors were prepared, with the proportion of kaons scaled up and down by 10\% for all \pt.
An analysis using these modified sets of priors was then performed and compared to the analysis with the standard priors. The corrected yields using the modified priors were divided by the yields using the standard set of priors for each Bayesian analysis method.
It was found that the average deviations from unity remained within roughly 2\% for each set of alternative priors with all of the Bayesian PID strategies. The smallest deviations were observed for the maximum probability and weighted methods, and for lower fixed probability thresholds.

In conclusion, since the corrected yield is largely independent of the choice of PID method, and considering the improvement in signal-to-background ratio and (at low $\pt$) significance, we conclude that the Bayesian PID method is a valid choice for the analysis of \DtoKpi. However, it must be noted that although the stricter Bayesian methods give a purer signal, they also create an increased reliance on the Monte Carlo simulations to correct for the lower efficiencies, as well as introducing a greater dependence on the choice of priors in those $\pt$\, intervals that have lower statistical significance.

\FloatBarrier

\section{Application of Bayesian PID for \Lc analysis}
\label{sec:lcres}

In this section we present an exploratory study of \LctopKpi  in pp collisions at $\sqrt{s}=7$\,\tev using a Bayesian PID approach. 
This decay channel suffers from a high level of combinatorial background, even in the relatively low-multiplicity environment of a pp collision. Due to the short decay length of the $\Lc$ ($\cTau=59.9\,\mim$~\cite{PDG}) and current limitations in the spatial resolution of the ITS, topological and geometrical selections alone are not sufficient to reduce this background and extract a signal.
This means that a robust PID strategy is required to select $\Lc$ candidates.

Here we compare analyses of $\LctopKpi$ using the $\nsigma$ approach, a Bayesian approach using the maximum-probability criterion, and an alternative ``minimum-$\sigma$'' approach that mimics the Bayesian strategy. This analysis used ${\sim}3\times10^8$ events from pp collisions at $\sqrts=7\,\TeV$ collected during \run{1}.

Both the TPC and TOF responses were used for the \nsigma approach. A 3$\sigma$ selection was applied on the TOF signal for kaons, pions and protons at all $\pt$, while in the TPC a $\pt$-dependent \nsigma selection (between 2 and 3$\sigma$) was applied for protons and kaons, and a 3$\sigma$ selection was applied for pions at all $\pt$. If the track had a valid TOF signal, only this was used for the PID; otherwise the TPC response was used.
For the Bayesian PID method, the TPC and TOF responses were combined; in cases where the TOF signal was not available, only the TPC response was used.

The minimum-$\sigma$ strategy serves as a middle ground between the $\nsigma$ analysis and the Bayesian approach. It is also based on the deviation from the expected detector signal, but in this case only the species resulting in the smallest \nsigma value is chosen for each track, making this an exclusive selection. The TPC and TOF responses are combined by adding their respective \nsigma values in quadrature. The potential benefit of this approach is that it uses similar logic to the Bayesian approach, without requiring the input of any priors.

In order to build $\Lc$ candidates, triplets of tracks were reconstructed and selected, based on PID and topological selections.  Pairs of oppositely charged tracks were reconstructed from the set of selected tracks, 
and a secondary (decay) vertex was computed. Selections were applied to each pair based on the distance of closest approach between the two tracks, and the distance between the primary and the secondary vertex of the reconstructed pair. 
A third track was then associated to each selected pair to form a triplet. The secondary vertex of the triplet was calculated and the 
$\Lc$ candidates were selected using:
\begin{itemize}
 \item the quality of the reconstructed vertex, which was determined using the quadratic sum of the distances of the single tracks from the calculated secondary vertex;
 \item the distance between the primary and the secondary vertex of the triplet (\rm{decay\ length}); and
 \item the cosine of the pointing angle. The pointing angle is the angle between the 
momentum vector of the reconstructed $\Lc$ candidate and the
$\Lc$ flight line reconstructed from the line joining the primary and secondary vertices.
\end{itemize}

\begin{figure}[h!]
\centering
\includegraphics[width=0.9\textwidth]{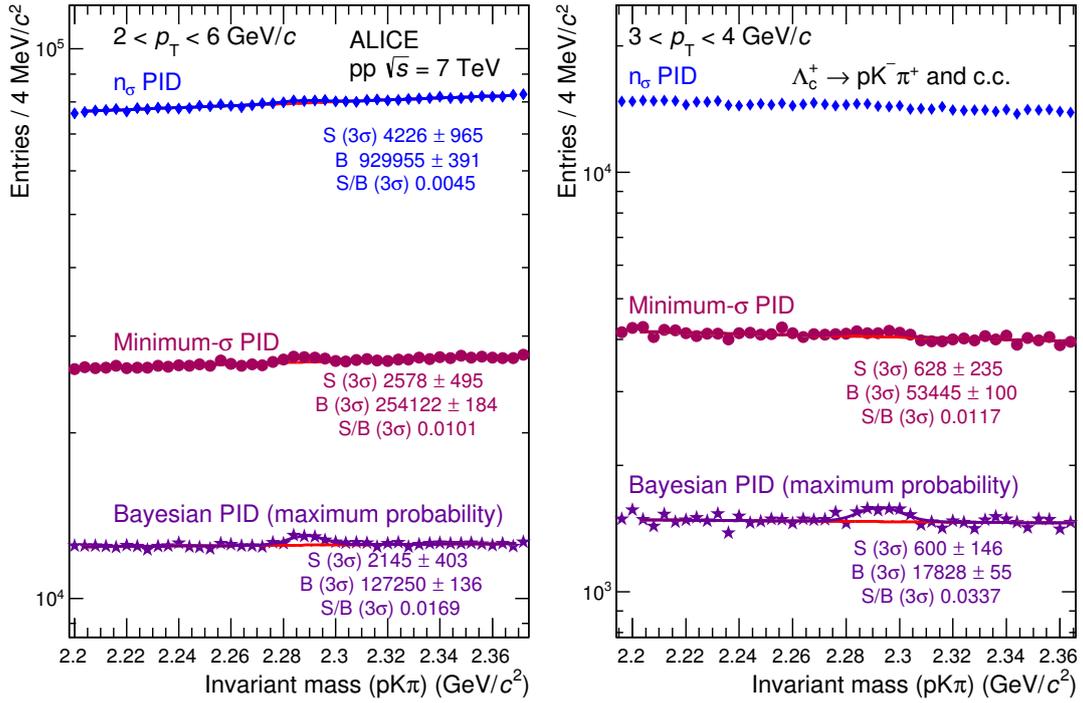}
\caption{Invariant mass spectra of $\LctopKpi$ using \nsigma PID, minimum-$\sigma$ PID and Bayesian PID for (left) $2<\pt<6$\,\gevc and (right) $3<\pt<4$\,\gevc. Due to the low statistical significance, it was not possible to extract a stable signal for \nsigma PID for $3<\pt<4\,\GeVc$, therefore this fit and its results are not shown.}
\label{fig:lc-pt-all}
\end{figure}
The invariant mass plots for pK$\pi$ candidates with $2<p_{\rm T}<6$\,\gevc are shown in the left-hand panel of \figref{fig:lc-pt-all} for the standard \nsigma PID approach (top), the alternative minimum-$\sigma$ approach (middle) and Bayesian PID (bottom). 
Over this \pt interval, the statistical significance determined from the yield extraction (as defined in \secref{sec:d0res}) is $4.4\pm1.0$ for \nsigma PID, $5.1\pm1.0$ for minimum-$\sigma$ PID and $6.0\pm1.1$ for Bayesian PID.
In addition, the signal-to-background ratio is three times higher in the Bayesian case than for \nsigma PID, with a background that is reduced by a factor of approximately seven. On the other hand, the Bayesian PID method yields fewer $\Lc$ candidates in the peak, indicating a lower selection efficiency.

The equivalent invariant mass distributions for $3<\pt<4$\,\gevc are shown in the right-hand panel of \figref{fig:lc-pt-all}. In this case, the \nsigma PID approach yields an insufficient statistical significance to extract a stable yield. As with the $\pt$-integrated case, the statistical significance improves when using the other two methods ($2.7\pm1.0$ for minimum-$\sigma$ and $4.4\pm1.1$ for Bayesian PID), and we again find a marked improvement in the signal-to-background ratio for these two approaches.
From this we conclude that the Bayesian approach represents the best candidate PID method for future measurements of the \Lc production cross section.

\FloatBarrier

\section{Conclusions and outlook}
\label{concl}

A Bayesian method for particle identification has been presented and validated for a variety of analyses.
A comparison between different PID selection methods in ALICE was performed,  with a focus on testing
the suitability of Bayesian PID techniques. A selection based on the Bayesian probability
can be interpreted as a request on the purity, and therefore the main benefit of these techniques is that they increase the purity of the extracted signal by combining information from different detectors via a relatively simple technique.

The iterative procedure used to extract the prior probabilities (corresponding to the relative particle abundances) was outlined for \pp collisions at $\sqrts=7\,\TeV$, \pPb collisions at $\sqrtsNN=5.02\,\TeV$ and \PbPb collisions at $\sqrtsNN=2.76\,\TeV$. In each collision system, and for every centrality and multiplicity class studied in \PbPb and \pPb collisions, the extracted priors were found to be consistent with the true particle abundances seen in data.

The ability of Monte Carlo simulations to compute efficiencies and misidentification probabilities was tested via high-purity samples of pions, kaons and protons
from the two-prong decays  $\Kzs \to \piMinus\piPlus$, $\rmLambda \to \proton\piMinus$, and $\phimes \to \Kminus\Kplus$ in \pPb collisions at $\sqrtsNN=5.02\,\TeV$.
Equivalent analyses were also performed using \nsigma PID. It was found that the variation in the result when using Bayesian PID as opposed to \nsigma PID was on the order of 5\% in each case. 

Comparisons between the Bayesian approach and PID methods used in measurements already published by ALICE in pp collisions at $\sqrts=7\,\TeV$ were performed for the analysis of the \pt-differential yields of identified pions, kaons and protons. In this case, the previous analysis used a combination of \nsigma~PID, unfolding methods and spectra obtained with kinks.
A simple maximum-probability Bayesian PID method was used in this analysis. When compared, the results from \nsigma PID and Bayesian PID were found to be consistent with one another within uncertainties.

A detailed comparison of different Bayesian PID strategies was performed for the analysis of \DtoKpi in pp collisions at $\sqrts=7\,\TeV$. Through the use of different probability thresholds, the trade-off between 
selecting a purer sample and having a stronger dependence on the description of the detectors in Monte Carlo simulations
was demonstrated. An increase in the signal-to-background ratio over the \nsigma PID method was seen for all \pt for all of the tested Bayesian strategies. The statistical significance was found to be similar or greater than \nsigma PID for all of the Bayesian methods other than a strict 80\% probability threshold. In every case, the corrected yield was found to be stable against the choice of PID method when comparing with the \nsigma PID method used for \Dzero mesons.
The dependence on the choice of priors was also investigated, and it was found that the uncertainty on the corrected yield due to the choice of priors was within ${\sim}2\%$. We conclude that the systematic uncertainties arising from the choice of Bayesian PID method are of a similar order to those of the \nsigma approach.

In summary, a good level of consistency is seen in both of the full analyses considered in Sections~\ref{sec:spectra} and \ref{sec:d0res} with respect to previously reported results. Because of this, we conclude that the validity of the Bayesian PID approach has been successfully assessed.

Furthermore, the improved performance of the Bayesian PID approach in certain scenarios will prove beneficial for future analyses. The case of the $\Lc$ baryon was presented as an example
where a simple \nsigma approach is unable to yield a stable signal, while the Bayesian approach is able to combine information from the different detectors effectively without the need to develop a complex variable-$\sigma$ selection. 
This presents a promising option for a more comprehensive study of the $\Lc$ production cross section in \pp and \pPb collisions.

The analyses presented here are currently limited to charged hadrons detected in the TPC and TOF detectors. However, the method can
be extended to other particle species and detectors within the ALICE central barrel, and we expect to further test and refine these techniques in future.
In particular, it will be possible to examine the performance of the Bayesian approach for electron identification with the TRD at full azimuthal coverage in \run{2}.

\newenvironment{acknowledgement}{\relax}{\relax}
\begin{acknowledgement}
\section*{Acknowledgements}

The ALICE Collaboration would like to thank all its engineers and technicians for their invaluable contributions to the construction of the experiment and the CERN accelerator teams for the outstanding performance of the LHC complex.
The ALICE Collaboration gratefully acknowledges the resources and support provided by all Grid centres and the Worldwide LHC Computing Grid (WLCG) collaboration.
The ALICE Collaboration acknowledges the following funding agencies for their support in building and
running the ALICE detector:
State Committee of Science,  World Federation of Scientists (WFS)
and Swiss Fonds Kidagan, Armenia;
Conselho Nacional de Desenvolvimento Cient\'{\i}fico e Tecnol\'{o}gico (CNPq), Financiadora de Estudos e Projetos (FINEP),
Funda\c{c}\~{a}o de Amparo \`{a} Pesquisa do Estado de S\~{a}o Paulo (FAPESP);
National Natural Science Foundation of China (NSFC), the Chinese Ministry of Education (CMOE)
and the Ministry of Science and Technology of China (MSTC);
Ministry of Education and Youth of the Czech Republic;
Danish Natural Science Research Council, the Carlsberg Foundation and the Danish National Research Foundation;
The European Research Council under the European Community's Seventh Framework Programme;
Helsinki Institute of Physics and the Academy of Finland;
French CNRS-IN2P3, the `Region Pays de Loire', `Region Alsace', `Region Auvergne' and CEA, France;
German Bundesministerium fur Bildung, Wissenschaft, Forschung und Technologie (BMBF) and the Helmholtz Association;
General Secretariat for Research and Technology, Ministry of Development, Greece;
National Research, Development and Innovation Office (NKFIH), Hungary;
Department of Atomic Energy and Department of Science and Technology of the Government of India;
Istituto Nazionale di Fisica Nucleare (INFN) and Centro Fermi -
Museo Storico della Fisica e Centro Studi e Ricerche ``Enrico Fermi'', Italy;
Japan Society for the Promotion of Science (JSPS) KAKENHI and MEXT, Japan;
Joint Institute for Nuclear Research, Dubna;
National Research Foundation of Korea (NRF);
Consejo Nacional de Cienca y Tecnologia (CONACYT), Direccion General de Asuntos del Personal Academico(DGAPA), M\'{e}xico, Amerique Latine Formation academique - 
European Commission~(ALFA-EC) and the EPLANET Program~(European Particle Physics Latin American Network);
Stichting voor Fundamenteel Onderzoek der Materie (FOM) and the Nederlandse Organisatie voor Wetenschappelijk Onderzoek (NWO), Netherlands;
Research Council of Norway (NFR);
National Science Centre, Poland;
Ministry of National Education/Institute for Atomic Physics and National Council of Scientific Research in Higher Education~(CNCSI-UEFISCDI), Romania;
Ministry of Education and Science of Russian Federation, Russian
Academy of Sciences, Russian Federal Agency of Atomic Energy,
Russian Federal Agency for Science and Innovations and The Russian
Foundation for Basic Research;
Ministry of Education of Slovakia;
Department of Science and Technology, South Africa;
Centro de Investigaciones Energeticas, Medioambientales y Tecnologicas (CIEMAT), E-Infrastructure shared between Europe and Latin America (EELA), 
Ministerio de Econom\'{i}a y Competitividad (MINECO) of Spain, Xunta de Galicia (Conseller\'{\i}a de Educaci\'{o}n),
Centro de Aplicaciones Tecnológicas y Desarrollo Nuclear (CEA\-DEN), Cubaenerg\'{\i}a, Cuba, and IAEA (International Atomic Energy Agency);
Swedish Research Council (VR) and Knut $\&$ Alice Wallenberg
Foundation (KAW);
Ukraine Ministry of Education and Science;
United Kingdom Science and Technology Facilities Council (STFC);
The United States Department of Energy, the United States National
Science Foundation, the State of Texas, and the State of Ohio;
Ministry of Science, Education and Sports of Croatia and  Unity through Knowledge Fund, Croatia;
Council of Scientific and Industrial Research (CSIR), New Delhi, India;
Pontificia Universidad Cat\'{o}lica del Per\'{u}.
\end{acknowledgement}

\bibliographystyle{utphys}
\bibliography{paper}

\newpage
\appendix
\section{The ALICE Collaboration}
\label{app:collab}

\bigskip 

J.~Adam$^{\rm 39}$, 
D.~Adamov\'{a}$^{\rm 84}$, 
M.M.~Aggarwal$^{\rm 88}$, 
G.~Aglieri Rinella$^{\rm 35}$, 
M.~Agnello$^{\rm 110}$, 
N.~Agrawal$^{\rm 47}$, 
Z.~Ahammed$^{\rm 132}$, 
S.~Ahmad$^{\rm 19}$, 
S.U.~Ahn$^{\rm 68}$, 
S.~Aiola$^{\rm 136}$
% ...

\bigskip

\bigskip 

\textbf{\Large Affiliation Notes}

\bigskip 

% ...
$^{\rm IV}$ Also at: M.V. Lomonosov Moscow State University, D.V. Skobeltsyn Institute of Nuclear, Physics, Moscow, Russia

\bigskip

\bigskip 

\textbf{\Large Collaboration Institutes}

\bigskip 

% ...
$^{19}$ Department of Physics, Aligarh Muslim University, Aligarh, India\\
% ...
$^{35}$ European Organization for Nuclear Research (CERN), Geneva, Switzerland\\
% ...
$^{39}$ Faculty of Nuclear Sciences and Physical Engineering, Czech Technical University in Prague, Prague, Czech Republic\\
% ...
$^{47}$ Indian Institute of Technology Bombay (IIT), Mumbai, India\\
% ...
$^{68}$ Korea Institute of Science and Technology Information, Daejeon, South Korea\\
% ...
$^{84}$ Nuclear Physics Institute, Academy of Sciences of the Czech Republic, \v{R}e\v{z} u Prahy, Czech Republic\\
% ...
$^{88}$ Physics Department, Panjab University, Chandigarh, India\\
% ...
$^{110}$ Sezione INFN, Turin, Italy\\
% ...
$^{132}$ Variable Energy Cyclotron Centre, Kolkata, India\\
% ...
$^{136}$ Yale University, New Haven, Connecticut, United States\\
% ...

\bigskip 


\end{document}